\documentclass{aa}  
\usepackage{epsfig}
\usepackage{graphicx}
\usepackage{xcolor}
\usepackage{txfonts}
\usepackage{amssymb}
\usepackage{amsfonts}
\usepackage{natbib}
\usepackage{hyperref}
\hypersetup{
  colorlinks=true, 
  urlcolor=blue, 
  linkcolor=blue, 
  citecolor=blue 
}

\bibpunct{(}{)}{;}{a}{}{,}

\begin{document}

\title{Characterisation of the stellar activity of M dwarfs. II. Relationship between Ca, H$\alpha$, and Na chromospheric emissions\thanks{Tables \ref{tab_targets} and \ref{tab_correl} are only available in electronic form at the CDS via anonymous ftp to cdsarc.u-strasbg.fr (130.79.128.5) or via http://cdsweb.u-strasbg.fr/cgi-bin/qcat?J/A+A/.} }

    \titlerunning{Characterisation of stellar activity of M dwarfs. Chromospheric emissions}

\author{N. Meunier \inst{1}, L. Mignon \inst{1,2} , Kretzschmar M. \inst{3}, X. Delfosse \inst{1} 
}
\authorrunning{N. Meunier et al.}

\institute{
Univ. Grenoble Alpes, CNRS, IPAG, F-38000 Grenoble, France \email{nadege.meunier@univ-grenoble-alpes.fr} 
\and 
Observatoire astronomique de l'Université de Genève, 51 chemin Pegasis, 1290 Versoix, Switzerland
\and
LPC2E, University of Orl\'eans, CNRS, 3A avenue de la Recherche Scientifique, 45071 Orl\'eans Cedex 2, France
}

   \date{}


\date{Received ; Accepted}

\abstract
{The chromospheric emission estimated in the core of different lines, such as Ca II H \& K, Na D1 and D2, and H$\alpha$, is not always correlated between lines. In particular, the Ca II H \& K and H$\alpha$ emission time series   are anti-correlated for a few percent of the stars, contrary to what is observed on the Sun. This puzzling result has been observed for both solar-type stars and M stars.  }  
{Our objective is to characterise these relationships in more detail using complementary criteria, and based on a large set of spectra obtained with HARPS for a large sample of M dwarfs. This should allow to evaluate whether or not additional processes are required to explain the observations.  }
{We analysed the time average and variability of the Ca, Na, and H$\alpha$ emissions for 177 M stars ranging from subspectral types M0 to M8, paying particular attention to their (anti-)correlations on both short and long timescales as well as slopes between indices. We also computed synthetic H$\alpha$ time series based on different assumptions of plage properties. We compared our findings with observations in order to evaluate whether or not the main observed properties could be reproduced. }
{The statistical properties of our sample, in terms of correlations and slopes between indices at different timescales, differ from what we previously obtained for FGK stars: there are fewer stars with a null correlation, and the correlations we find show  a weaker
dependence on timescale. However, there can be a large dispersion from one season to another for stars with a well identified low or negative correlation. We also specify the complex relationship between the average activity levels, with a clear indication of a change in the sign of the slope from the relation between Ca and H$\alpha$ (and between Na and H$\alpha$) for the most massive M dwarfs. In addition, we observe a change in slope in the Na--Ca relation at an intermediate activity level. At this stage, we are not able to find simple plage properties that, alone, are sufficient to reproduce the observations. However, the simulations already allow us to point out that it is not straightforward to compare the temporal variability correlation and the integrated indices. Our findings also demonstrate the need for complex activity patterns to explain some of the observations.}
{We conclude that the relation between the three indices examined here exhibits a large diversity in behaviour over the sample studied. More detailed simulations with complex activity patterns are necessary to understand these observations. This  will teach us about plage properties for this type of star. }


\maketitle
%

\section{Introduction}

A large fraction of M dwarfs are active stars characterised by a strong chromospheric emission with a dependence on spectral type \cite[e.g.][]{reiners10,reiners12}, or with a photometric variability interpreted as being due to the presence of spots of plages \cite[e.g.][]{suarez16}. This suggests a dynamo process at play in many of those stars, including examples in the fully convective regime \cite[][]{chabrier06,browning08,yadav15,yadav16,wright16}. This transition between partially convective and fully convective occurs around spectral type M3.5 or a mass of around 0.35~M$_{\odot}$ \cite[][]{chabrier97}. The relationships between the luminosity in X and the rotation period \cite[e.g.][]{pizzolato03,wright11} and between the chromospheric emission in Ca and the rotation period \cite[e.g.][]{astudillo17,wright18} are other important diagnoses of their variability: two regimes are observed, a saturated regime for fast rotators (average activity level independent of rotation) and a linear regime for slow rotators, typically above 10 days. In addition, long-term cycles have been observed in both regimes and for both partially and fully convective stars \cite[e.g.][]{buccino11,suarez16,ibanez19,ibanez19b,ibanez20,mignon21c}. This has been studied  over limited samples  \cite[][]{gomes11,gomes12,robertson13} and with larger samples, mostly in photometry or chromospheric emission
 \cite[][]{savanov12,robertson13,vida13,vida14,suarez16,kuker19,mignon21c}, and sometimes in combination with a spectroscopic analysis \cite[][]{suarez18}.
 Stellar activity also affects exoplanet studies, impacting the detectability of low-mass exoplanets around the above types of stars \cite[see ][for a review on processes]{meunier21b,meunier22c}, and potentially influencing the habitability of the exoplanets \cite[e.g.][]{ip00,buccino06,buccino07,vonbloh07,segura18,vidotto22}.

 Studies of the chromospheric emission  focused on typically three indices, the emission in the Ca II lines \cite[e.g.][]{astudillo17,suarez18}, in the Na doublet D$_1$ and D$_2$ \cite[e.g.][]{diaz07,gomes12}, and in H$\alpha$ \cite[e.g.][]{robertson13}, either individually,  combined \cite[e.g.][]{ibanez23,gomes11}, or combined with the Ca  
infrared triplet (IRT) observations \citep{lafarga21}. \cite{gomes11}  also considered the He I line. These different indices have often been used while assuming that they were providing equivalent information on the activity level. This is particularly important for M dwarfs, because the usual $\log R'_{HK}$ indicator, which is based on the Ca II lines, is noisier than for FGK stars due to the low flux emitted in the UV continuum by these stars. This means that the H$\alpha$ emission is an interesting alternative as an activity proxy, because it should allow a much better signal-to-noise ratio. 
 However, these emissions do not form at the same height in the atmosphere. 
 The Na lines form in the lower chromosphere \cite[][]{mauas94,andretta97,fontenla16}, the Ca II H \& K lines in the middle chromosphere \cite[][]{mauas94} to the upper chromosphere \cite[][]{mauas00,houdebine09,fontenla16}, and the H$\alpha$ line in the upper chromosphere \cite[][]{mauas94,fuhrmeister05,houdebine09,fontenla16,leenaarts12}.  The relationship between these indicators then proved to be complex, mostly from two points of view.

 A first issue concerns the relationship between averaged (in time) activity indicators for a sample of stars. Although they are globally correlated over a stellar sample (when taking temperature effects into account), non-linearity between the Ca and H$\alpha$ emission is expected for M dwarfs at low activity levels. Given the fact that when activity increases, the absorption in H$\alpha$ first increases before switching to emission \cite[see also][]{stauffer86,cram87},  \cite{cram79} proposed that this should theoretically lead to a U-shape of the Ca II--H$\alpha$ relationship. 
 Many attempts have been made to observe this effect in M stars, first with non-simultaneous observations of Ca II and H$\alpha$ \cite[][]{giampapa89,robinson90,strassmeier90} and then on simultaneous observations. On the other hand, a strong dispersion was also observed \cite[][]{houdebine97,gomes11,frasca16}. These latter authors found a flat relation at low Ca levels, usually with some large dispersion in H$\alpha$, followed by an increase above a certain threshold. \cite{rauscher06,walkowicz09} observed hints of the effect in a sample of M stars, with a U-shaped lower envelope in the relation between the equivalent width in H$\alpha$ and Ca. The U-shape was more clearly observed by \cite{scandariato17}, with a negative slope in the low-activity regime,  albeit with a limited sample of stars.
 \cite{robinson90} suggested that the Ca emission is due to a strong heterogeneity in the coverage of the stellar surface by magnetic field (i.e. solar-like), while the H$\alpha$ emission is less variable because it is less sensitive to local temperature and density, and could therefore be associated with a more homogeneous chromospheric appearance \cite[see also][]{rauscher06}.
  \cite{walkowicz09} suggested that this large dispersion is intrinsic and proposed several qualitative scenarios concerning the importance of plages on these stars, with different relative importance of active regions compared to quiet chromosphere.    
 It is therefore important to study this effect in a much larger sample and to check whether or not there is a dependence on spectral type.

A second issue was raised by \cite{cin07}, who found that a few percent of the stars in their sample of 101 F-M stars exhibited an anti-correlation between the Ca and H$\alpha$ time series instead of a positive correlation. This was later confirmed with a large sample of FGK stars \cite[][]{gomes14,meunier22} and for M dwarfs \cite[][]{gomes11}, although all stars in the sample of \cite{ibanez23} were found to exhibit a positive or close-to-zero correlation. A few individual studies also reported these effects. For example, \cite{robertson13} found that GJ~581 exhibit an anticorrelation between the Na and the H$\alpha$ emission, while \cite{dimaio20} observed an anticorrelation between Ca and H$\alpha$ for AD Leo, with a relationship that may depend on timescale. \cite{walkowicz09} pointed out a large dispersion between Ca and H$\alpha$ for this latter star, despite the fact that it lies in the active regime where a correlation is expected. Furthermore, our analysis of long-term variability in a large sample of M dwarfs \cite[][hereafter Paper I]{mignon21c}  showed that the periodicities found in the different indices are not always the same. The Ca variability is, for example, in some cases dominated by a very long trend not seen in H$\alpha$ and dominated by a shorter variability. The possibility to interpret these anticorrelations by the presence of filaments exhibiting some absorption in H$\alpha$ was proposed by \cite{meunier09a} and explored \cite[][hereafter Paper II]{meunier22} for FGK stars. \cite{gomes22} also recently proposed that anticorrelation could be sensitive to the bandpass used to compute the H$\alpha$ index for FGK stars.

The objective of the present paper is therefore to address these two issues by studying a  large sample of 177 stars. We computed and analysed the three activity indices in Ca II, Na, and H$\alpha$ for this sample. We thus extend the work done for FGK stars in Paper II to this large sample of M stars. This large sample allows us to focus on the dependence on spectral type, and to study both averaged indices and variability at different timescales, which provide complementary information. 
The outline of this paper is as follows. In Sect.~2, we present the stellar sample, and the chromospheric emission indices (Ca, Na, and H$\alpha$) used in the analysis and in the selection process on the data. In Sect.~3, we briefly focus on the study of averaged indices, while Sect.~4 presents a detailed analysis of the relationships between indices on different temporal scales. Synthetic series based on simple plage models are analysed in Sect.~5. We conclude in Sect.~6.

\section{Data and timescales}

We first describe the stellar sample and the HARPS data used in the analysis, and then the selection process and the activity indices. We define the seasons we use to study the relation between the different indices on short  (typically below 150 days) and long  (above 150 days) timescales in Sect.~4 and 5. More details can be found in Paper I. 

\subsection{Stellar sample and data}
 
We retrieved all identified HARPS observations of M stars from the ESO archive (Spectral Data Products Query Form\footnote{\url{http://archive.eso.org/wdb/wdb/adp/phase3_spectral/form}}), as in \cite{mignon21a,mignon21c} until 02-2020. These observations come from many programs of M dwarfs (listed in the acknowledgements). The sample of stars we study in this paper are the M dwarfs of the solar neighbourhood, which we selected from different catalogues:   \cite{gaidos14}, \cite{winters2014solar}, \cite{winters21},   \cite{stauffer2010accurate}, and \cite{henry18}.
After a selection process described below (Sect.~2.3),
for stars with multiple measurements on a given night, the indices are averaged to give a single measurement per night. We considered stars with at least ten nights of observations and a long enough temporal coverage to be able to implement a study of long-term variability as in Paper I (i.e. stars with a temporal coverage shorter than six times the rotation period are removed), which led to a sample of 177 stars, which are listed in Table~\ref{tab_targets}.

\subsection{Activity indices}

We computed chromospheric activity indices as in Paper I, following \cite{gomes11} and \cite{astudillo17}. This was done for the Ca II H \& K lines, the Na doublet D1 and D2, and the H$\alpha$ line. This led to S indices for each of them, noted in the following $S_{\rm Ca}$, $S_{\rm Na}$, and $S_{\rm H\alpha}$, corresponding to the integrated flux in the lines divided by the continuum levels. For the Ca II H \& K lines, the integration in the core of the lines is performed with a triangular window, while it is performed with a simple square widow for Na and H$\alpha$ as usual \cite[see][for more details]{mignon21c}. The bandwidth is 0.5~\AA$\:$ and 1.6~\AA$\:$ for Na and H$\alpha$ respectively. Uncertainties on each measurement were computed as the quadratic sum of the flux photon noise and the read-out noise as in \cite{astudillo17} and \cite{mignon21c}.
Since we are mostly interested in variability, the S indices are sufficient to compare the variability between the three elements. 
The time-average indices were also computed, and these are discussed more specifically in Sect.~3. 

Finally, for comparison of average activity levels between all spectral types, we also computed the $\log R'_{HK}$ from $S_{\rm Ca}$, following the calibration by \cite{astudillo17},  and including a correction to ensure consistency with the Mount Wilson program with the law they derived from the comparison with \cite{wright04},
\begin{equation}
S_{M.W.} = 1.053 · S_{HARPS} + 0.026,
\end{equation}
where $S_{HARPS}$ was obtained from the HARPS spectra, and $S_{M.W.}$ is the corrected value.

 We chose a large H$\alpha$ bandwidth as in previous works. However, \cite{gomes22} found that this choice appears to impact the correlation between the H$\alpha$ index and the S-index for F-G-K stars. We therefore tested the lower bandwidth they used, 0.6~$\AA$, on a small sub-sample of stars representative of the different configurations.

\subsection{Selection process}\label{select}

We briefly describe our selection process, following Paper I. A first step of selection consisted in removing bad quality spectra based on the signal to noise ratio (hereafter S/N), using a fixed threshold (1 on the S/N in the order of Calcium lines and 10 on the S/N in the median order, $\sim$ 550 nm). Examples of spectra with low and high S/N are shown in Fig.~\ref{exsnr}. We then eliminated spectra with S/N values much lower than for the other observations of the star. Stars with a poor S/N but a very large number of observations during the corresponding night were kept however, since they were later averaged over the night, but these are very rare (it concerns only five stars in the final sample, and in most cases for one night only).

The second step of selection focuses on outliers and flares. The analysis was based on the distribution of values from a given time series: the gap between the bulk of the data and outliers was estimated, and uncertainties were taken into account to rejected those outliers (both highest values and/or lowest values, often corresponding to problem on the spectra). We note that the highest values could be due to either problems in the spectra, or strong flares, because we do not aim to characterise these events.
Importantly, all measurements rejected from the time series obtained for a given activity index were also rejected for the other two activity indices, to keep the same temporal sampling for the three  indices.

\begin{figure}
\includegraphics{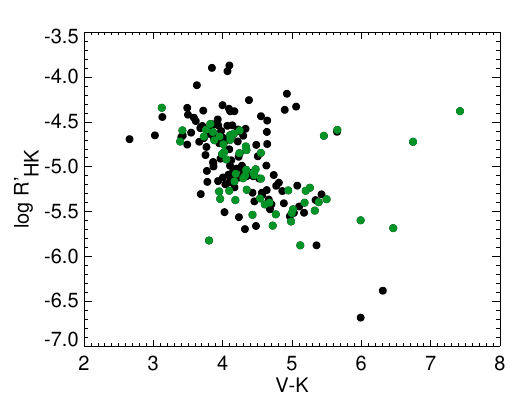}
\caption{$\log R'_{HK}$ versus V-K for the 177 stars in our nightly sample. Stars in green correspond to the subsample of 60 stars with at least four seasons of observations. 
}
\label{sample}
\end{figure}

\subsection{Global, short-term, and long-term analyses}
\label{sec24}

The study of the variability and of the correlation between indices was mostly based on the  time series as described above (one point per night, although there are naturally not an observation every night) for the sample of 177 stars, with more than ten nights of observations. Such time series include both short-term and long-term contributions of stellar activity.

Following Papers I and II, we also defined seasons for a subsample of stars with a good  temporal sampling, to apply statistical tests on those binned time series and then focus on the long-term (hereafter LT) and short-term (hereafter ST) variability separately as in Paper II: this is studied in Sect.~4.3. As in Paper I, seasons are defined as independent bins of 150 days, to average as well as possible the rotational modulation \cite[the rotation periods are longer than for FGK stars,][]{newton16} with at least five observations in the season, and gaps between observations inside the bin shorter than 40 days. The subsample includes stars with at least four seasons only, which leads to 60 stars: the number of seasons is then between 4 and 12, with a median of 6. 37 stars have at least 6 seasons and 14 more than 8. In summary, the short-term variability corresponds to variability observed within seasons, while long-term variability deals with variability from season to season.

\section{Average activity indices}
\label{sec3}

In this section, we first focus on time-averaged indices for each star, to be able to put the variability study in context. We first present and discuss the average activity level. Then  we consider relationships between averaged (over time) indices in the different  bins in T$_{\rm eff}$. 

\subsection{Basal flux in the Ca II chromospheric emission}

\begin{figure}
\includegraphics{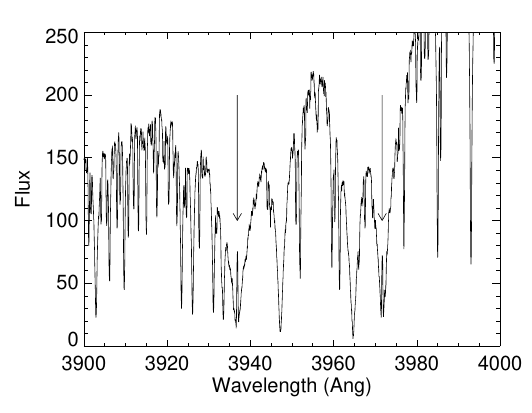}
\caption{Binned spectrum for the quiet star GJ~191 ($\log R'_{HK}$=-5.823). The two arrows highlight the existence of  an emission in the Ca II H \& K lines. }
\label{exsp}
\end{figure}

 We wish to evaluate if the quietest stars in our sample correspond to the basal flux (defined in this paper as the activity level due to acoustic heating, with no magnetic contribution) and if those stars still exhibit some kind of variability. In the latter case, all stars are of interest for the study of the correlations between activity indices.
Figure~\ref{sample} shows the time-averaged $\log R'_{HK}$ of the stars in our sample versus V-K. There is a decreasing trend, with a lower envelope decreasing with increasing V-K. As already seen in Paper II and \cite{astudillo17}, this trend is opposite to that of K stars, for which there is an increasing lower envelope \cite[see also][]{mittag13,borosaikia18,meunier18a,meunier19}. The lower envelope corresponds to the least active stars in the sample. We point out that this lower envelope does not  correspond to the basal flux  corresponding to no activity, because the spectra for stars at the level of this lower envelope still exhibits some emission in the core of the Ca II H \& K lines, even if it is weak, as illustrated for GJ~191 in Fig.~\ref{exsp} ($\log R'_{HK}$=-5.823, close to the lower envelope in Fig.~\ref{sample}). In fact, some of these very "quiet" stars also exhibit some significant variability despite their very low level of emission, such as GJ~191 for example, although with smaller amplitudes compared to more active stars.

\subsection{Relationship between averaged indices in Ca, Na and H$\alpha$}

Before focusing on temporal variability, we study the time-averaged indices for the sample of 177 stars, to be able to place the variability study in a broader context.  This is very useful when comparing simulated time series, since those should reproduce not only the observed variability, but also the average properties. The S-index computations do not include any normalisation related to temperature effects.  \cite{martinezarnaiz11}, \cite{scandariato17} and \cite{dimaio20}  normalised the spectra by subtracting the spectrum from a reference star assumed to be quiet and close in spectral type. However, these reference stars are not quiet, in the sense that there is some emission in the core of the Ca II H \& K lines and temporal variability\footnote{GJ205, GJ393, GJ273, and GJ4092 used in cite{martinezarnaiz11} are very active and variable \cite[][]{mignon21c,gomes11,suarez16}.}.
Since we also found  in the previous section that stars at the level of the lower envelope in $\log R'_{HK}$ still show an emission in the core of the Ca II H \& K lines, we  consider that such a normalisation is not reliable. 
Finally, \cite{walkowicz09} subtracted the photospheric spectrum from a model, but used the same model for all stars since they have all the same spectral type (M3), which may be a limitation. It is beyond the scope of this paper to build a new way to normalise these indices. 
We therefore consider here   T$_{\rm eff}$ bins  of 100 K, inside which we consider that there is sufficient homogeneity to look at the global relationship between averaged indices. To be able to constitute the most complete list of stars per bin, we estimated T$_{\rm eff}$ for the 10 stars for which we did not have a value from the literature, by fitting a linear law between T$_{\rm eff}$ and V-K for all other stars in our sample and then using that law to estimate their T$_{\rm eff}$.

\begin{figure}
\includegraphics{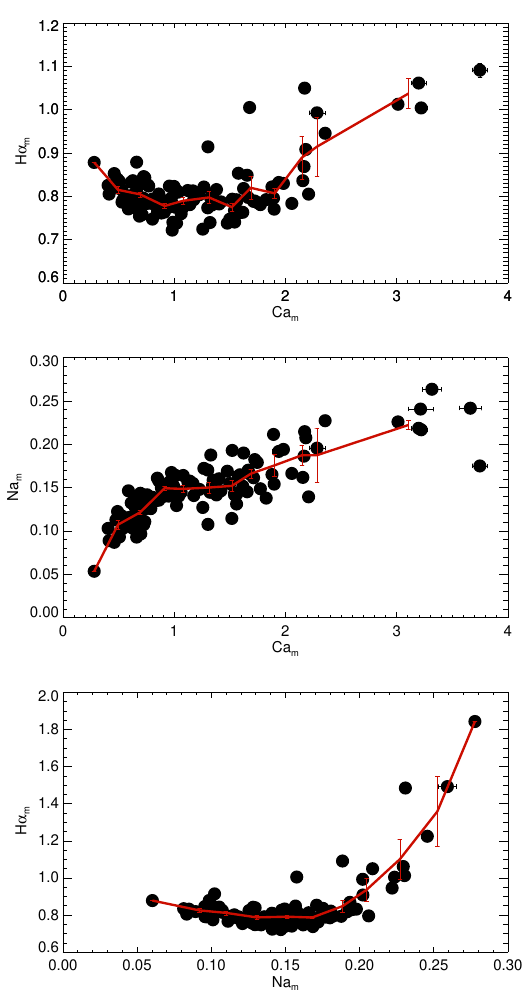}
\caption{
Relation between average indices ---without the most active stars--- for the three pairs of indices (from top to bottom) and the 3400-3900 K range. The values in ordinates have been shifted vertically to take the temperature effect into account (We selected points with Ca$_{\rm m}$ around 0.9 (for the two first plots) or Na$_{\rm m}$ around 0.15 (for the last plot), fitted a linear function to the index versus T$_{\rm eff}$ and applied the resulting correction to all points.). The red points and lines correspond to averages in bins in abscissa. 
}
\label{ex_moy}
\end{figure}

\begin{figure}
\includegraphics{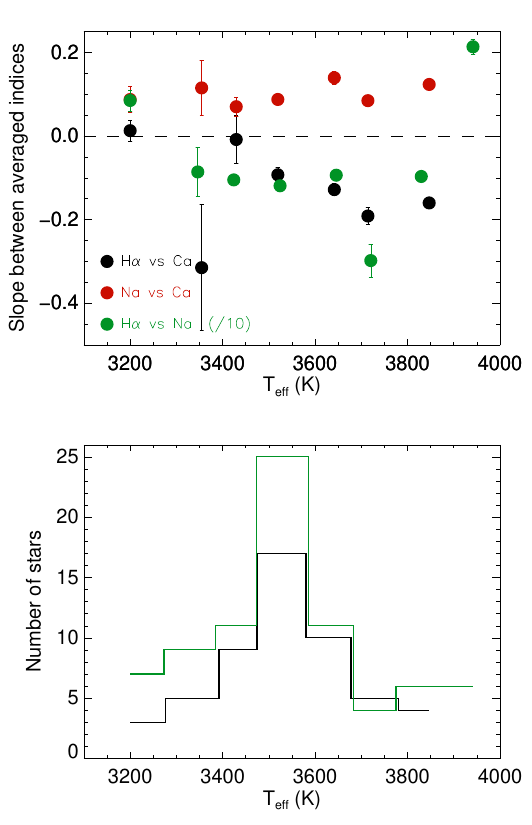}
\caption{Slopes between averaged indices for stars in the low-activity regime versus T$_{\rm eff}$ (upper panel), and the number of stars in each T$_{\rm eff}$ bin (lower panel), corresponding to the three pairs of indices. Slopes correspond to H$\alpha$ versus Ca (black), Na versus Ca (red), and H$\alpha$ versus Na (green, divided by 10 for clarity). The number of star per bin in black corresponds to the  H$\alpha$-Ca and Na-Ca pairs, while the number of stars in green corresponds to the H$\alpha$-Na pair. 
}
\label{slopemoy}
\end{figure}

In the following, the time-averaged $S_{\rm Ca}$, $S_{\rm Na}$ and $S_{\rm H\alpha}$ for each star are noted Ca$_{\rm m}$, Na$_{\rm m}$ and H$\alpha_{\rm m}$.
The relationships between them are shown in Appendix~\ref{appC} for all T$_{\rm eff}$ bins, for the whole range of activity levels, with a close-up on quiet stars. 
Despite a global positive correlation (computed as the Pearson coefficient) between averaged indices when considering all stars, 
H$\alpha_{\rm m}$ is fairly constant, and even decrease when Ca increases for the most quiet stars for certain T$_{\rm eff}$ bins and in the quiet star regime. 

The flat regime in the Ca$_{\rm m}$-H$\alpha_{\rm m}$ relation was observed in the previous works cited above, usually associated with a large dispersion. The U-shape of the lower envelope was also observed by \cite{rauscher06} and \cite{walkowicz09}. The negative slope was already observed by \cite{scandariato17}, based on a few stars only. We confirm this behaviour on a much larger sample. This seems to be consistent with the prediction of \cite{cram79}, with an increased H$\alpha$ absorption when the Ca emission increases, before going into emission at higher activity levels. This interpretation is discussed in Sect.~5.  The effect is also seen when considering H$\alpha_{\rm m}$ versus Na$_{\rm m}$, but not for the Ca-Na pair, although we observe a change in slope for  Ca$_{\rm m}$ around 0.9.

 This behaviour is  seen in the 3400-3900 K range only (with 113 stars in that T$_{\rm eff}$ range), as illustrated in Fig.~\ref{ex_moy}, for the relatively quiet stars. 
To quantify this effect, we computed the slope in the low activity regime, that is for Ca$_{\rm m}$<0.9 \footnote{The 0.9 threshold in Ca$_{\rm m}$ 
roughly corresponds to a threshold in $\log R'_{HK}$ of around -5, although there is an overlap between the two distributions corresponding to stars below and above the 0.9  threshold. The $\log R'_{HK}$ values have been normalised to the Mount Wilson ones following the law provided in \cite{astudillo17}.}  for the two first pairs of indices, and for Na$_{\rm m}$ below 0.15 for the last one.  All slopes are shown versus T$_{\rm eff}$ in Fig.~\ref{slopemoy}. The change is sign is not observed below 3400-3500~K, nor with mass bins below $\sim$0.4 M$_{\odot}$, suggesting that the non-linear relationship between Ca (or Na) and H$\alpha$ is present only in partially convective stars.

\section{Relationship between activity index time series}
\label{sec4}

\begin{figure}
\includegraphics{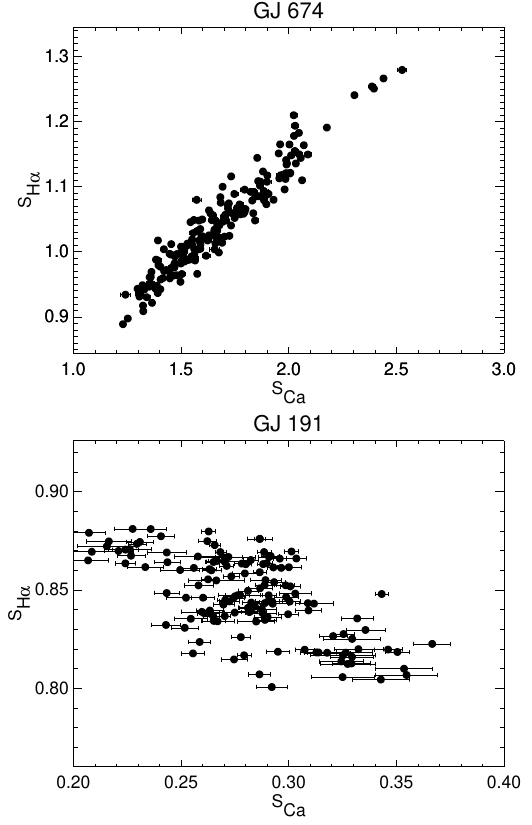}
\caption{H$\alpha$ versus Ca indices for GJ 674 (upper panel) and GJ 191 (lower panel). H$\alpha$ uncertainties (method described in Sect.~2.2) are hardly visible because they are much smaller than the Ca uncertainties. }
\label{exemple0}
\end{figure}

\begin{figure*}
\includegraphics{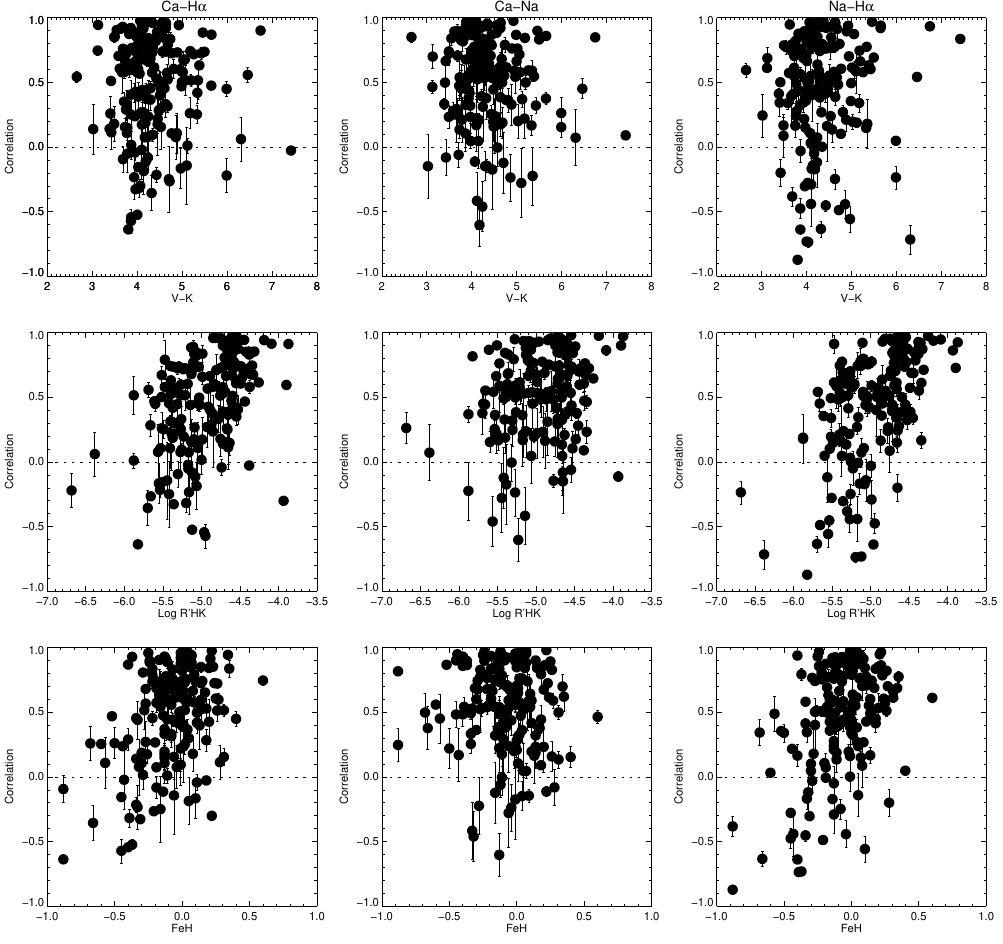}
\caption{Global correlation between pairs of indices versus V-K (upper  panels),  $\log R'_{HK}$ (middle panels), and metallicity (lower panels) for the 177 stars in our sample, for the three different correlations (from left to right): Ca--H$\alpha$, Ca--Na, and Na--H$\alpha$.}
\label{correl}
\end{figure*}

This section is devoted to the detailed analysis of  the observed variability in Ca, Na and H$\alpha$ based on various criteria, and more specifically on their relationship. We mostly study the correlation between time-series, and the slope obtained from a linear fit between time-series to quantify the relationship between activity indicators. We also compare these criteria at different timescales. The analysis was performed on the three time series for each star, 
$S_{\rm Ca}$, $S_{\rm Na}$, and $S_{\rm H\alpha}$, corresponding to the three pairs of indices.

\subsection{Examples of relationships between time series}

We first illustrate the diversity in the behaviours that are observed in our large sample with a few typical examples that are representative of the different configurations.
They are  are shown in Fig.~\ref{exemple1} and Fig.~\ref{exemple2}. For each star, we first show the three time series, corresponding respectively to Ca, Na, and H$\alpha$. Then, we show the relationships for the three corresponding pairs. 
This illustrates the wide range of observed configurations. Some stars exhibit a very good correlation (GJ~588, GJ~674) or moderate correlation (GJ~9592, GJ~273) between all indices, while for others stars some indices are correlated (most often Ca-Na, for example GJ~191 or GJ~832) while other indices are anti-correlated (GJ~191) or not correlated (GJ~832). Finally, GJ~406 exhibits a particular behaviour, with a correlation between Na and H$\alpha$, while Ca is not correlated with either of those indices. The correlation is very poor for all pairs of indices for GJ~3470.
We study these correlations in more details in the next section, for the whole sample. 

The two main stars with a strong Ca-H$\alpha$ anticorrelation are  GJ~191, followed by GJ~87: There is a striking difference between the behaviour of those  stars compared to a star with a very good correlation (Fig.~\ref{exemple0}).  
 We confirm the anticorrrelation found for GJ~581 by \cite{robertson13}, but not the poor correlation obtained by \cite{dimaio20} or \cite{buccino14} for GJ~388 (even without observations corresponding to flares), which in our case is very well correlated. The sample studied in \cite{ibanez23} includes 13 stars which are also in our sample: there is a qualitative agreement between the different observed behaviour, the main exception being  GJ~388 as well. 
We recall that during the selection process (Sect.~2.3), we eliminated the strongest flares. However, medium size flares may still impact in the time series and degrade the correlation, especially for the most active fast rotators \cite{ibanez23}. 

We tested the impact of using a narrow bandwidth in H$\alpha$ such as in \cite{gomes22} on a few stars that are representative of the different configurations we observe. We found that they do not show any significant difference. In particular, the strong anticorrelation observed for GJ~191 and GJ~87 remains similar.

\subsection{Global variability}
\label{sec42}

In this section, we analyse the correlations (defined as the Pearson coefficient), denoted C, between the time series for the three pairs of indices of our sample of 177 stars and compare them. The correlations are shown in Table~\ref{tab_correl} for all stars. We first analyse how these correlations depend on stellar properties, and then study their distributions, their dependence on timescale, and the relationship between the different correlations. The correlation between indices are then compared with the parameters of a  linear fit between indices.

\subsubsection{Dependence of the correlations on stellar properties}

\begin{figure}
\includegraphics{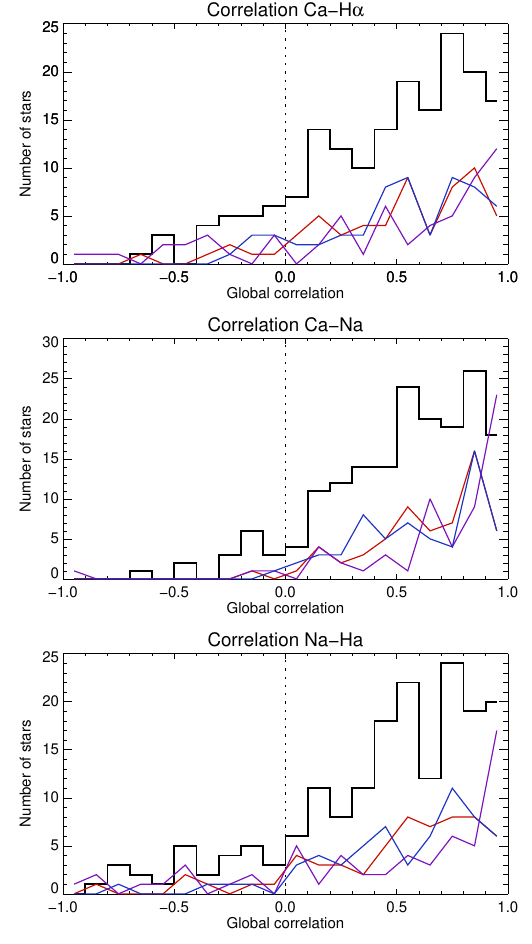}
\caption{Distribution of the global correlations for the whole sample of 177 stars (thick black lines) for the three pairs of indices (from top to bottom). Distributions in colour are for the subsample of stars with a sufficient number of seasons covered by our data. We show global correlations (red), long-term correlations (purple), and short-term correlations (blue).}
\label{distcorrel}
\end{figure}

Given the large diversity of configurations, we first study how they depend on stellar parameters. This is particularly interesting for M dwarfs, which exhibit two regimes, the fully convective and partially convective ones. 
None of the correlations exhibits any trend with V-K (upper panels of Fig.~\ref{correl}) nor with T$_{\rm eff}$. 
\cite{ibanez23} did not observe any such correlation either on their sample of 29 M dwarfs.

On the other hand,  a moderate correlation is observed between C and the average activity level characterised  by the average  $\log R'_{HK}$ (middle panels of Fig.~\ref{correl})
We observe in both cases an increasing lower envelope, which is not unlike what we observed for FGK stars in Paper II, although the transition was much sharper for FGK stars. We also observe a lack of stars with very strong correlations (above 0.9) for the quietest star. \cite{ibanez23} found no strong dependence of the Ca-H$\alpha$ correlation on $\log R'_{HK}$ on their sample of 29 stars, however their less correlated star is also the less active. may be due to the fact that the stars with a very strong correlation between indices tend to be stars that are dominated by the rotational modulation and a strong short-term variability, which may not be detectable in very quiet stars. The main exception to the general pattern is GJ~406, already mentioned above, which is a very low mass and active star with very weak correlations between indices. 
Our sample includes very few very low mass stars and therefore it is difficult to generalise. 

Finally, we analyse the relationship between  C and metallicity, for the 170 stars for which we found an estimate in the literature. \cite{scandariato17} indeed already mentioned a weak correlation between the two, but found that it was not significant enough to conclude. 
There is a moderate correlation for the pairs with H$\alpha$, similar to the correlation with $\log R'_{HK}$ (lower panels of Fig.~\ref{correl}). 
The interpretation of this significant correlation with metallicity is complex however, because there is also a relationship between the $\log R'_{HK}$ and FeH (see Appendix~\ref{appE} for more details: the cause could be an age-metallicity relation (AMR)  among the stars in the solar vicinity, the oldest stars being less active because their rotation has slowed down. Several publications suggesting that such a correlation was indeed present in the disc of the Milky Way, with the oldest stars being statistically the least metallic \cite[][]{twarog80,rochapinto00,soubiran08}. But more recent studies show that this question is still highly debated, with \cite{haywood13}, \cite{bergemann14} or \cite{rebassa16} concluding that the AMR is very noisy or even absent for stars in the galactic disk with ages between 0 and 7 Gyr. The interpretation of these disagreements is often attributed to the difficulty of determining precise stellar ages.
It is therefore not easy to disentangle the different effects. 
Furthermore, as shown in Appendix~\ref{appC}, the relationship between activity, correlations with H$\alpha$ and FeH seems to be present for quiet stars only, for T$_{\rm eff}$ above 3400~K (no difference is seen for active stars), which corresponds to a threshold similar to the change in slope in Sect.~3. The trend is also opposite to what is observed for FGK stars \cite[][]{jenkins08,meunier07}. This points towards different dominant processes controling the correlations for M and FGK stars.

We conclude that the correlations involving H$\alpha$ differs from the Ca-Na relation, as for the time-averaged studies in Sect.~3, with a sensitivity on the average activity level.

\begin{figure*}
\includegraphics{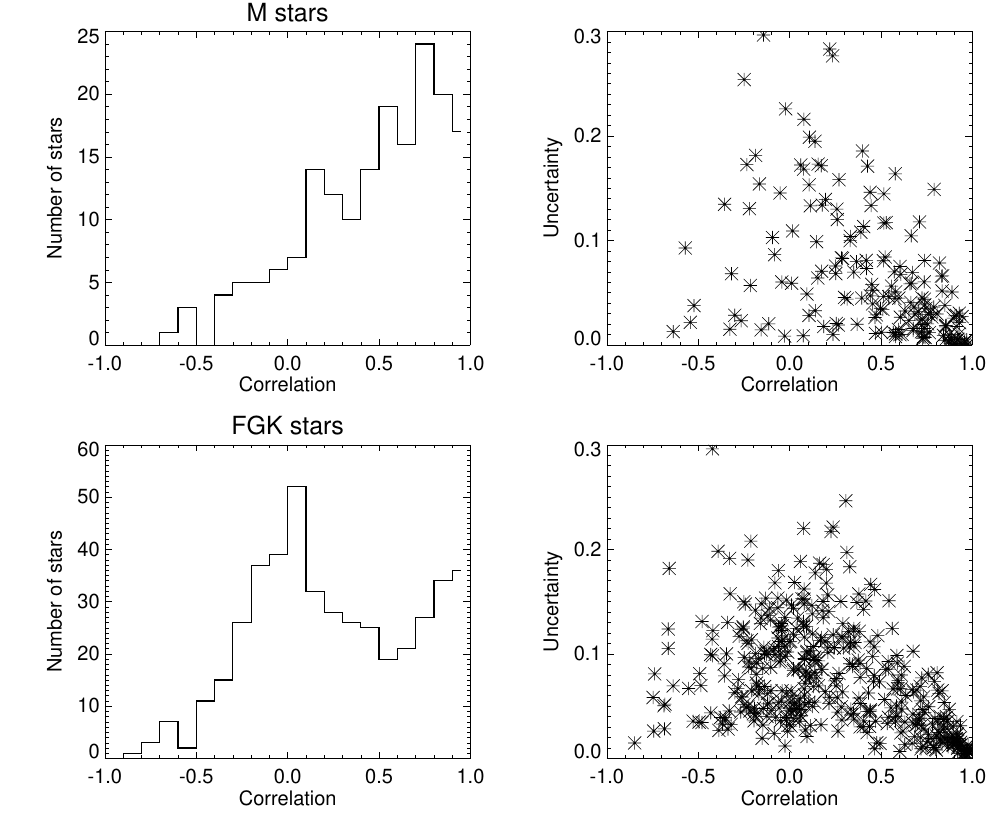}
\caption{Distribution of the Ca--H$\alpha$ correlations for the 177 M stars (upper panel, this paper) and for the sample of 441 FGK stars in \cite{meunier22}. The right panels show the uncertainty on the correlation versus the correlation. 
}
\label{compfgk}
\end{figure*}

\begin{table}
\caption{Global correlation rates}
\label{tab_glob}
\begin{center}
\renewcommand{\footnoterule}{}  
\begin{tabular}{lllll}
\hline
Selection & FGK & \multicolumn{3}{c}{M} \\
          &  Ca-H$\alpha$  &  Ca-H$\alpha$  &  Ca-Na & Na-H$\alpha$ \\
\hline
C$>$0.5 & 31.1 & 54.2 & 60.5 & 54.8 \\
C$>$0.3 & 42.6 & 67.8 & 76.3 & 71.2 \\
C$<$-0.5 & 2.9 & 2.3 & 0.6 & 4.0 \\
C$<$-0.3 & 8.8 & 4.5 & 1.7 & 7.9 \\
$C\neq 0$ (1$\sigma$) & 81.9 & 89.3 & 92.7 & 94.4 \\
$C\neq 0$ (3$\sigma$) & 55.8 & 79.7 & 75.7 & 85.3 \\
|C| $<$0.2  & 36.3 & 18.1 & 13.6 &  14.1 \\
\hline
\end{tabular}
\end{center}
\tablefoot{Percentage of stars in different correlation (C) regimes for the sample of 441 FGK stars in \cite{meunier22} and for the 177 M stars in the present paper. 
}
\end{table}

\subsubsection{Distribution of the correlations}

We first compare the correlations of the different pairs through their distribution, in particular to establish if the anticorrelations are seen only with H$\alpha$. These are also a useful reference for comparison with the distribution of correlations computed at different timescales. 
The curves in black in Fig.~\ref{distcorrel} show this distribution of the C values for the three pairs of indices for the whole sample of 177 stars. They are globally similar, with many more stars in the correlated regime, with few stars strongly anticorrelated. We recall that the choice of the H$\alpha$ bandwidth does not significantly impact those measurements.   Those exhibit small uncertainties\footnote{Uncertainties on the Pearson correlations are computed with a Monte-Carlo approach to propagate the uncertainties on individual measurements on the final correlation.}  for the pairs with H$\alpha$, and are therefore significant. For the Ca-Na anticorrelations however, fewer stars are in this regime, and they exhibit high uncertainties, so that they may be compatible with no correlation rather than representing a significant anticorrelation.
Furthermore, there are many less stars (about twice less, Table~\ref{tab_glob}) with a correlation between indices close to zero compared  to FGK stars (Paper II), for which there was clearly a peak in the distribution of the H$\alpha$-Ca correlation around zero, corresponding to a very specific behaviour (Fig.~\ref{compfgk}).

\subsubsection{Relations between correlations}

\begin{figure}
\includegraphics{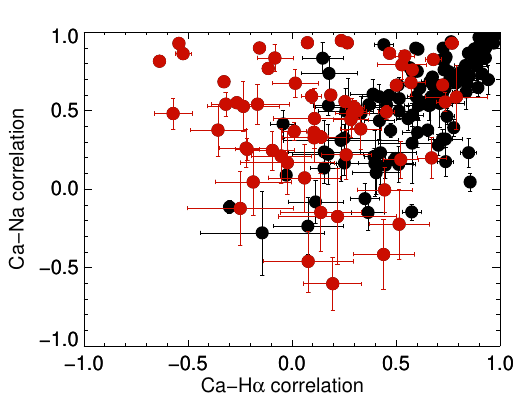}
\includegraphics[width=9cm]{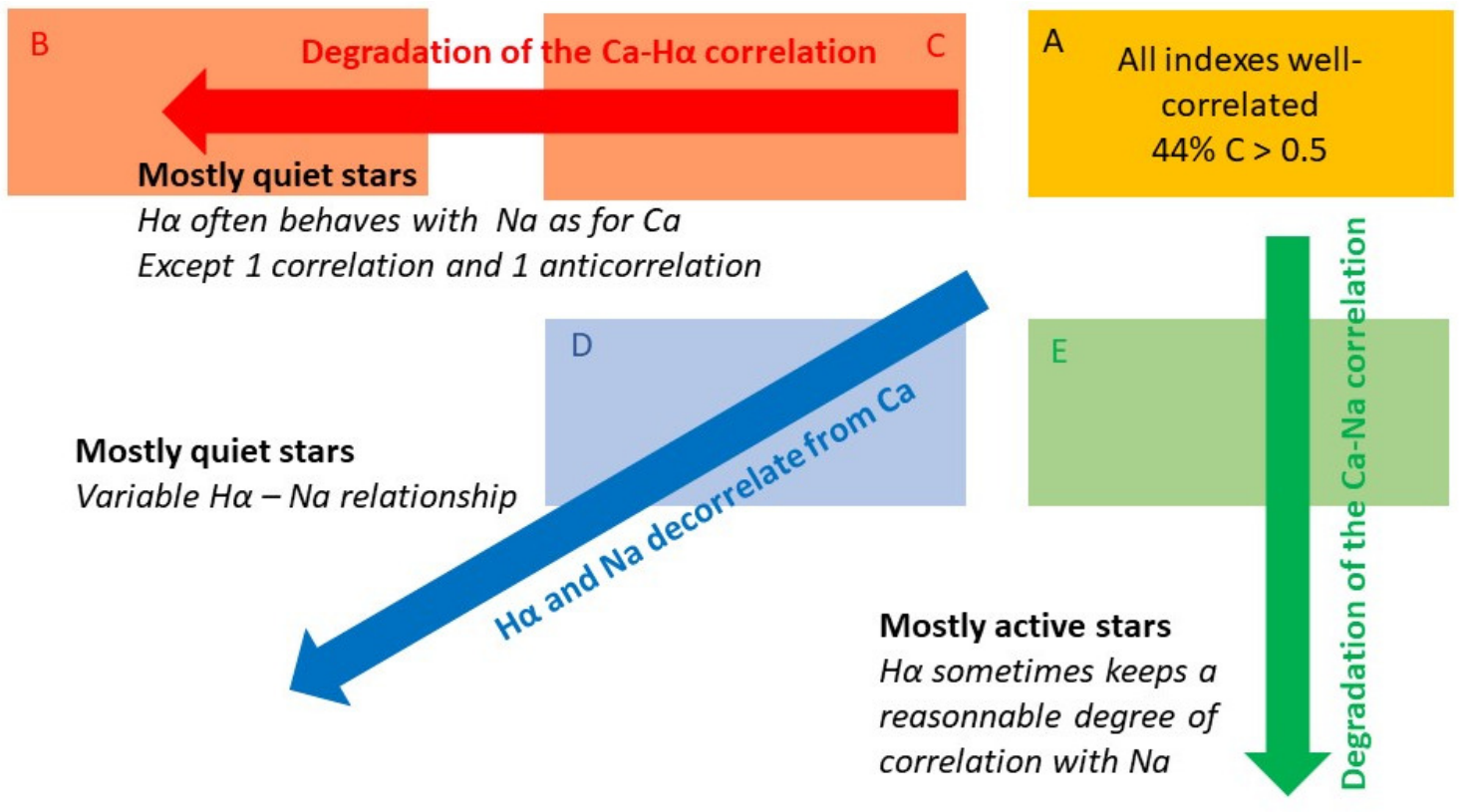}
\caption{Ca--Na correlations versus Ca--H$\alpha$ correlations (upper panel) for quiet stars (red) and active stars (black). The lower panel shows a schematic representation of the relationship between the different correlations (the different categories are discussed in the main text). 
}
\label{relatcorrel}
\end{figure}

We now analyse how the correlations relate to each other in more details and how these relations depend on activity level. The first panel in Fig.~\ref{relatcorrel} shows the Ca-Na correlation versus the H$\alpha$-Ca correlation, where the quietest stars are indicated in red. There is a large amount of stars in the upper right corner, which corresponds to stars for which all activity indices are well correlated. 
For the other stars, we observe a large dispersion and a large diversity, which is schematised in the diagram shown in the second panel, with three directions, starting from this regime with very good correlations:
\begin{itemize}
\item{Stars conserving a good Ca-Na correlation, but with a degraded Ca-H$\alpha$ correlation (red arrow). Those are mostly quiet stars, which is coherent with the relationship between this correlation and the activity level seen above. The strongly anticorrelated (in Ca-H$\alpha$) stars in category B (5 stars), and those with this correlation close to zero in category C (11 stars). Examples are illustrated in Fig.~\ref{exemple2} and the most representative stars are GJ~191, GJ~87 (category B), and GJ~667C GJ~832 GJ~3341 GJ~433 (H$\alpha$ usually not correlated with Na),  GJ~3634 with a good Na-H$\alpha$ correlation, and GJ~3804 for which H$\alpha$ anticorrelates with Na (category C).}
\item{Stars conserving a good Ca-H$\alpha$ correlation, with a degraded Ca-Na correlation (green arrow). Those are mostly active or moderately active stars. 6 stars have $|C|<$0.2 for Ca-Na (category E), some examples being 2MASSJ19301369-5456193 (H$\alpha$ not correlated with Na), and GJ~43, GJ~639 (with a Na-H$\alpha$ anticorrelation but a large uncertainty). 
}
\item{Stars for which both Ca-H$\alpha$ and Ca-Na are degraded (blue arrow). There are very few cases (7 stars with both $|C|<$0.2), the most prominent one being GJ406 (H$\alpha$ remains relatively correlated to Na), already mentioned above, and GJ~2049, GJ~9201 (H$\alpha$ not correlated with Na). Apart from GJ~406, they are usually quiet stars (category D)  that are not in the saturated regime: the low rotation rate may lead to a variability with a worse signal-to-noise ratio, hence a correlation closer to zero. In the case of FGK stars, a large fraction of the stars in this regime corresponded to well-defined correlations, despite a usually low average $\log R'_{HK}$ with a significant Ca variability (they correspond to stars with a correlation close to zero and low uncertainties in Fig.~\ref{compfgk}). Such a population appears to be absent in our M dwarf sample. } 
\end{itemize}

The general panorama is therefore complex, with a large diversity of configurations when considering the three pairs of indices. In addition, the degradation of the Ca-H$\alpha$ and Ca-Na do not correspond to stars with similar activity levels, and may therefore involve different processes.

\subsubsection{Relationship between correlations and slopes}

\begin{figure*}
\includegraphics{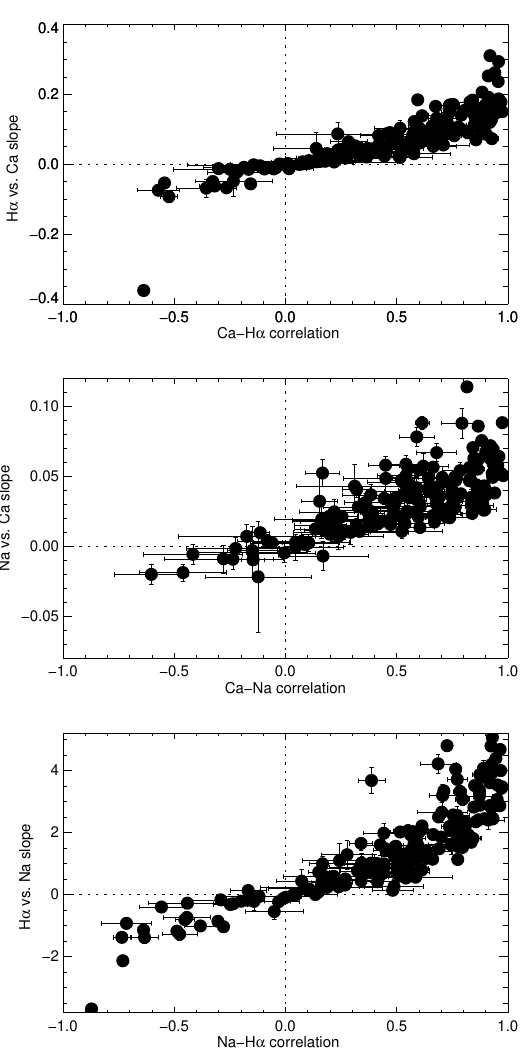}
\includegraphics{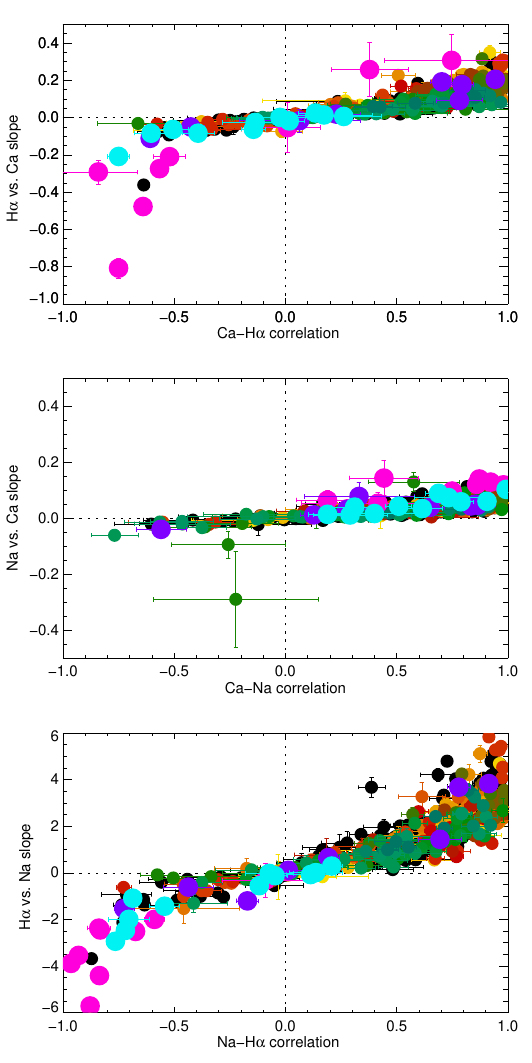}
\caption{Slope between indices versus correlation for the different pairs of indices (from top to bottom). We show global quantities for all stars (left) and with superposition of the  ST slopes for individual  seasons (right), for stars for which we have data covering at least six seasons. The ST slopes of a given star all have the same colour. A few stars are indicated with a larger symbol and are discussed in the text: GJ~191 (pink), GJ~87 (purple), and GJ~581 (light blue). 
}
\label{slope1}
\end{figure*}

We now analyse the slopes between the activity indices for each star as a complementary criterion, and compare them with the global slopes based on average values (Sect.~3.2): We expect them to have the same sign than the correlation,  but they also contain information about the respective amplitude of the variability. This comparison is particularly useful when comparing with simulations in Sect.~5.  We use the following 
linear fit, for example for the Ca-H$\alpha$ pair for a given star:
\begin{equation}
\label{eqlin}
    {\rm S_{H\alpha}}(t)=S_{ind} \times {\rm S_{Ca}}(t) + k
\end{equation}
where $S_{ind}$ is the slope and $k$ an offset.
The slopes are shown in Table~\ref{tab_correl} for all stars. 
Figure~\ref{slope1} (left panels) shows the slope between pairs of indices versus the global correlation. There is a strong, expected relationship between the two. When considering stars with a good  positive correlation, there is some dispersion, which is mostly due to temperature effects (related to different variability amplitudes, see Sect.~3): $S_{ind}$ is indeed directly affected by these effects, while the correlation is not. A similar plot for FGK stars from the data in Paper II would be similar. However, we note that the most anticorrelated star in our sample, GJ~191 (lower left corner of the Ca-H$\alpha$ and Na-H$\alpha$ plots) departs from the general trend, as it corresponds to a much stronger H$\alpha$ variability than the other stars (about a factor 3), relatively to the Ca or Na variability. There is no such drop for the Ca-Na plot, which shows few stars in the lower-left corner of these plots.

\subsubsection{Relationship between  slopes and average index behaviour}

\begin{figure}
\includegraphics{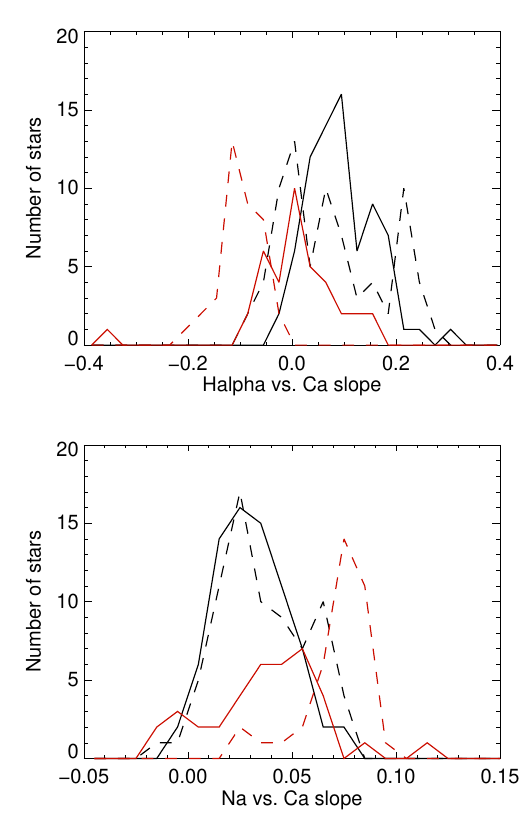}
\caption{
Distribution of the slopes $S_{ind}$ (solid lines; slope on the stellar time series defined in Sect.~\~4.2.4) and  $S_{av}$ (dashed lines; slope on average indices defined in Sect.~\~4.2.5), for Ca versus H$\alpha$ (upper panel) and Ca versus Na (lower panel). We show only the values for the 3500-3900K range, for Ca$_{\rm m}$<0.9 (red) and Ca$_{\rm m}$>0.9 (black). }
\label{lienmoy_na}
\end{figure}

In this section, we compare the behaviour derived from the average indices in Sect.~3, with those derived from the variability of each star based on the various indices studied in Sect.~4 to understand what controls those slopes and correlations. We wish to check whether the two are similar, or if their relationship is more complex. Indeed, we found that although the stars exhibiting an anticorrelation, that is with a negative $S_{ind}$ (Eq.~\ref{eqlin}, Sect.~4.2.4).  were in the quiet star regime (in which there was negative slope for the average H$\alpha$ versus Ca or Na in the 3400-3900 K range, Fig.~\ref{slopemoy}) some stars in this low activity regime do have positive $S_{ind}$ values.  Using the result of Sect.~3, we attribute a value of this slope, hereafter denoted $S_{av}$, to each star, to compare with $S_{ind}$.

We first focus on the relationship between H$\alpha$ and Ca.
Figure~\ref{lienmoy_na} (upper panel) shows the distribution of the two slopes for stars in the 3500-3900 K range in two activity regimes. 
In the high activity regime (Ca$_{\rm m}$>0.9), all $S_{ind}$ values are positive, which is compatible with $S_{av}$. 
However, for the 37 stars in the quiet activity regime (red curves),  we  observe  $S_{ind}$ with both signs, despite the fact that we are in a regime with a negative $S_{av}$ (red dashed curve). 
Even if $|S_{av}|$ is high, the individual variation in H$\alpha$ over time may be more flat, and sometimes reversed.
We conclude that even if both the negative $S_{av}$ and the existence of stars with anticorrelation between Ca and H$\alpha$ suggest that this could be explained by the properties of the non-monotonous variation of the H$\alpha$ emission \cite{cram79}, their relationship is sufficiently complex and not unequivocal, so that a better understanding is necessary.

For the Na-Ca relationship, $S_{ind}$ and $S_{av}$ are both mostly positive. 
The two distributions, shown separately in the low and high activity regimes (lower panel in Fig.~\ref{lienmoy_na}) are very different for $S_{av}$, but they are not for $S_{ind}$. The change in slope that we noticed in Sect.~3 is  not seen for $S_{ind}$: it is therefore unlikely to be due to a property that varies over time, but rather to a property intrinsic to the star, for example its fundamental parameters.

\subsection{Long-term and short-term variability}

\begin{figure*}
\includegraphics{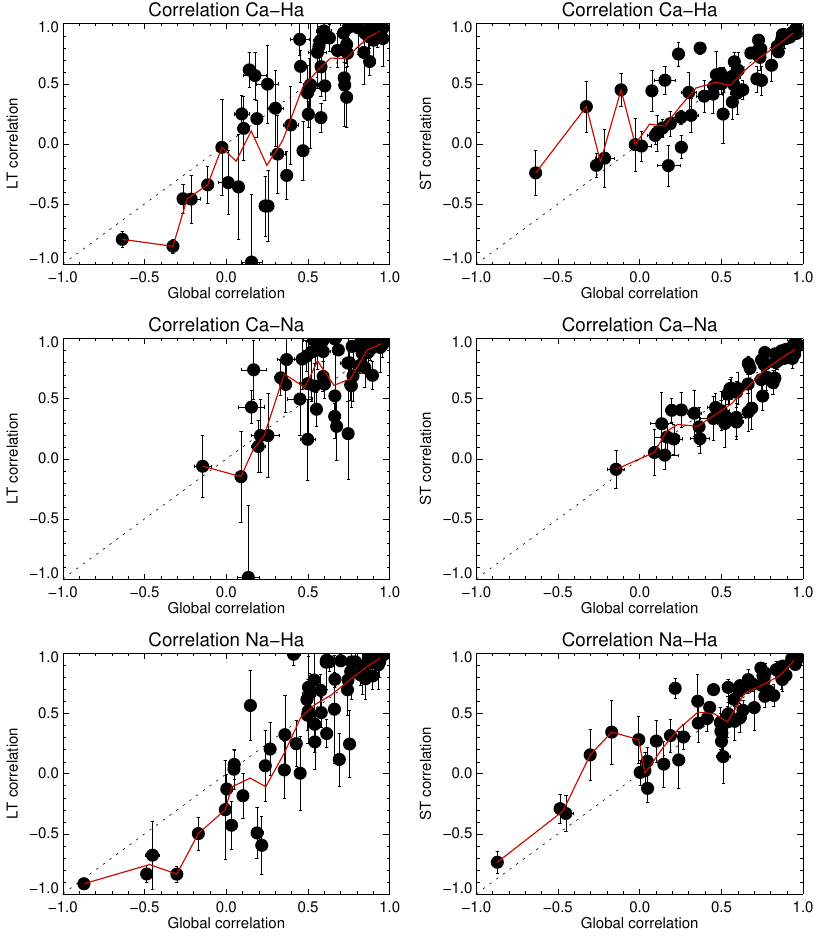}
\caption{LT (left panels) and ST (right panels) correlations versus global correlation for the three pairs of indices (from top to bottom). The red line indicates the average LT or ST, respectively, averaged in bins of global correlations. The dotted line is the y=x line.  }
\label{correlltct}
\end{figure*}

\begin{figure}
\includegraphics{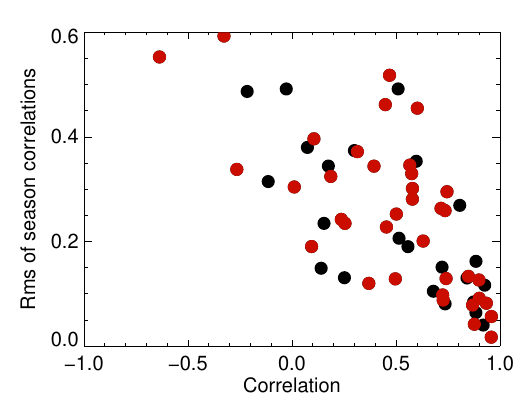}
\caption{Root mean square between ST Ca--H$\alpha$ correlations versus the global correlation for stars for which our data cover at least four seasons. Points in red correspond to stars for which our data cover at least six seasons.
}
\label{rmsltct}
\end{figure}

We now consider how the correlations and slopes behave at different temporal scales, on a subsample of 60 stars for which we defined at least four seasons, allowing to derive ST (typically below 150 d)  and LT (above 150 d) variabilities, as defined in Sect.~2.4. This approach proved to be very useful for FGK stars (Paper II) because it allowed to identify that scenarii based on plages only had difficulties to reconcile the observations at all timescales simultaneously. 

\subsubsection{Distributions of the correlations }

As in Sect.~4.2.2 for the global correlations, a first step is to compare the distributions of the correlations
Figure~\ref{distcorrel} shows the distributions of the correlations for the three pairs of indices at long and short timescales. The Kolmogorov-Smirnov probability between those distributions shows that there is in general no significant difference between them (which is again very different from FGK stars), the only exception being the LT Na-Ca correlation, which exhibits a small deficit in low correlations.

\subsubsection{Comparison with the global correlations}

We directly compare the LT and ST correlations with the global ones in Fig.~\ref{correlltct}. 
These are in general in good agreement, with the exception of the anticorrelations, with closer to -1 in the LT case compared to the global correlations, and closer to 0 in the ST case. This is not observed for the positive correlation, conversely to the FGK stars.
 The Ca-Na LT and ST correlations are also  very similar on average to the global correlation, although there is a significant dispersion, in particular for the LT correlations.  
As a consequence, the comparison between the correlations at different timescales is not very discriminant, except for the few anticorrelated stars. 

\subsubsection{Variability over different seasons}

We now analyse how the ST correlations vary over time. The ST variability should be strongly dominated by rotation. If the structures responsible for the variability have a lifetime that is not smaller than the rotation period for example, the modulation therefore corresponds to the same structures over time and not to a variation of the actual activity level at the surface of the star (conversely to what could happen for LT variability, which could also be due to some diffuse network structures with no impact on ST variability).
Figure~\ref{rmsltct} shows the rms between the ST Ca-H$\alpha$ correlations of each star versus the global correlation. Plots for the other pairs of indices are similar. We observe a larger dispersion in rms toward lower global correlations: This means that for a given star, some seasons can be well correlated, while other can be poorly correlated or even anticorrelated, suggesting different activity pattern configurations over time, especially anticorrelated or poorly correlated stars. This is illustrated in the right panels of Fig.~\ref{slope1}: 
In general, the ST points lie along the global values, and  the points appear to lie in different regimes depending on the season. 
If the anticorrelations are due to plages properties in relation with the activity level as suggested by \cite{cram79} and the relationship between averaged indices (Sect.~3), we then expect the ST correlations to be related to the average activity level of the stars. We therefore performed tests to estimate the slope between the ST correlations and the average activity level for each of the stars, but the results are unfortunately not significant. Apart for possibly GJ~667C, the slope is never significantly different from zero, and it is  not possible  to conclude.

\subsubsection{Observed H$\alpha$ variability compared to Ca II and Na at different timescales}
\label{sec434}

The last property we study from the observed time series is the variability at different timescales for the different indices, which are then compared. The variability is defined here as the rms of the indices, either globally, during a season (ST), or between the season-averaged values (LT). This proved to be an important criteria when analysing the relationship between the Ca and H$\alpha$ variabilities in Paper II.

\begin{figure}
\includegraphics{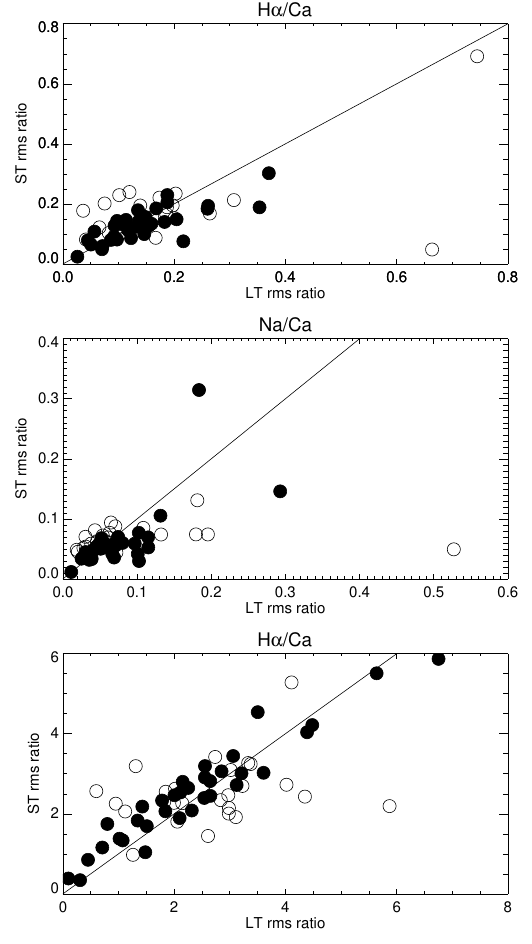}
\caption{Ratio between the ST rms and the LT rms for the three pairs of indices (from top to bottom). Filled circles correspond to the most variable stars (i.e. with a better S/N), the threshold is a global Ca rms higher than 0.1. The solid line is the y=x line.  One outlier in the middle panel (at a LT ratio of $\sim$1) is not shown here for clarity. 
}
\label{ratiorms}
\end{figure}

To compare the ST and LT variabilities, we computed the ratio between the H$\alpha$ and Ca rms at short timescale (i.e. over each season, which are then averaged over the seasons of the star), and compare it with the same ratio from the LT time series. The same computation is performed for the other pairs of indices. The results are shown in Fig.~\ref{ratiorms}. The relationship between the H$\alpha$-Ca ratio at the two timescales follows the y=x line, with some dispersion: This is in contrast with the FGK results in Paper II, for which there were many stars significantly above this line, with an excess of H$\alpha$ emission at ST compared to LT variability, relatively to the Ca variability, which could not be explained by the presence of noise.  The same is true for the H$\alpha$ to Na ratio. On the other hand, the Na-Ca relation shows a small lack of Na emission at short timescales compared to long timescales, relatively to the Ca variability, which remains to be understood.

We note that we observe a significant variability on short timescales, even in the low activity regime, which questions the interpretation by \cite{robinson90} that the flat regime by the fact that the H$\alpha$ emission was spread over the whole surface: The periodogram analysis performed in Paper I shows that among the stars in the low activity regime (3400-3900 K range), we found 23 stars with a peak a low period above the 1\% false alarm probability level, suggesting rotational modulation: 8 of these stars have a negative slope, but 15 have a positive slope despite the fact that they are in this regime (see discussion in Sect.~4.2.5).

\section{Synthetic correlations and slopes  based on a local emission law in plages}

Our objective is now to build models to test which conditions allow to retrieve the main properties of the observations found in the previous section. We follow two approaches. The first one is similar to Paper II: We build synthetic H$\alpha$ and Na time series from the Ca observed one. This approach is described in Appendix~\ref{appF}: It failed to reproduce moderate correlations, so that a more complex model is necessary. Also, conversely to the FGK analysis, ST and LT analysis does not allow to discriminate between assumptions. 

In the second approach, described here, we build several toy models with very simple configurations and increasing complexity. They are based on several parameters relating the emission in plages  for Ca and H$\alpha$,  based on a local law following the general prescription of \cite{cram79} for H$\alpha$.
The objective is not to reproduce all the complexity in this preliminary analysis, but to better understand  the impact of the different parameters affecting the measurements and confront them with observations. 
We first discuss these parameters, and then present results from simple models corresponding to different assumptions on these parameters. These models are compared with some of our observations to identify what is well reproduced and what is missing.

\subsection{General considerations}
\label{sec51}

It has been suggested that the expected relation between H$\alpha$ and Ca emissions from \cite{cram79} could explain the observed relation between averaged indices, which we observe in the 3500-3900 K range, as well as possibly stars with a negative correlation and slope. However, it is important to point out that the causal effect is not necessarily direct. Indeed, the prediction concerns the emission in one point of the atmosphere (either a quiet region, or plage), or in other words for a homogeneous atmosphere, while the observations correspond to the integrated emission over the disc. We know however that the atmosphere is heterogeneous, since emission spectra can not be reproduced with a single component \cite[e.g.][]{houdebine10b,houdebine11}. We also observe some modulation at the rotational timescales (Paper I) in all indices, including H$\alpha$, suggesting the presence of structures at the surface. We can therefore expect that the process of averaging over the disc  may lead to specific effects. In particular, we observed that the slope between indices for a given star is different (and often even the sign) from the slope derived from the averaged indices. In addition, we found stars in the flat regime (identified from the averaged index analysis) that are significantly  variable.

An example of effect occurring during the averaging process is the following \cite[see also the discussions in][]{walkowicz09}. Two main parameters can impact the averaged indices: the filling factor (hereafter ff) of plages (or structures in general), and the local properties as defined by the local relationship from \cite{cram79} for example. A small ff corresponding to a strong emission could correspond to a relatively low averaged emission but a positive correlation between indices. On the other hand, a large ff (corresponding for example to many very small structures, plages or similar to the solar network) and a weak emission, could be in the H$\alpha$ absorption regime, leading to an anticorrelation, while both configurations correspond to a similar average activity level defined by the  Ca level (in the following, reference to the activity level of a star corresponds to the Ca level). Another effect is the role of the basal flux (defined here as the flux with no magnetic field), corresponding to stars which would have not magnetic activity at all. However, the rest of the atmosphere (outside plages) is not necessarily quiet either, as shown in the solar atmosphere, with many network structures (by analogy with the solar case). This also includes some Ca flux in the quiet Sun \cite[][]{meunier18a}, which also shows some cycle-related variability. This should therefore affect the position of the stars in the plot  such as the one shown in Sect.~3 (Fig.~\ref{ex_moy}).

\subsection{General set-up}

\begin{figure}
\includegraphics{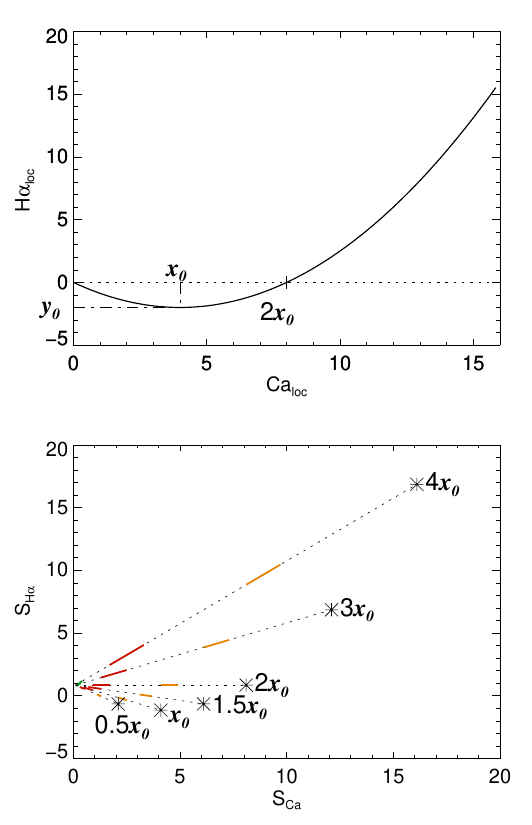}
\caption{Illustration of the law from Eq.~\ref{eqcram} (upper panel). The lower panel illustrates the range covered by $S_{\rm H\alpha}(t)$ versus $S_{\rm Ca}(t)$ for different Ca$_{\rm loc}$ (corresponding to different slopes) and different ff ranges: 0.5-0.6 (orange), 0.1-0.2 (red), and 0.01-0.02 (green), in a model with ${\rm Ca_{qs}}$=0. The black stars correspond to ff=1.  
}
\label{ca_ha_schema}
\end{figure}

In the following, we describe the variation of the local H$\alpha$ emission (H$\alpha_{\rm loc}$) as a function of the local Ca emission (Ca$_{\rm loc}$) following the general prescription of \cite{cram79}, that is first an absorption and then a emission when the Ca level is high enough. We did not find any quantitative prescription in the literature, we therefore use a simple second degree polynomial law to represent a decrease followed by an increase, as follows:
\begin{equation}
\label{eqcram}
    {\rm H\alpha_{loc}} = a {\rm Ca_{loc}}^2+b {\rm Ca_{loc}}
\end{equation}
This law therefore goes through a minimum (maximum absorption in H$\alpha$) for a Ca level of $x_0$, for which the H$\alpha$ emission is $y_0$ (negative value), and then increases again and becomes positive for a Ca level of 2$x_0$. With these notations, a=-$y_0/x_0^2$ et b=2$\times y_0$/$x_0$. This law can be applied to localised structures such as plages or to the other components of the atmosphere (expected to have a low level in Ca emission compared to plages). An illustration of the law is shown in the upper panel of Fig.~\ref{ca_ha_schema}.

We then considered the following  general model for the indices we computed from observed spectra at a given time: 
\begin{equation}
\label{eqtest1_ca}
S_{\rm Ca}= {\rm Ca_0} + {\rm Ca_{qs}} \times (1-{\rm ff}) + {\rm Ca_{pl}} \times {\rm ff}
\end{equation}
\begin{equation}
\label{eqtest1_ha}
S_{\rm H\alpha}= {\rm  H\alpha_0} + {\rm H\alpha_{qs}} \times (1-{\rm ff}) + {\rm H\alpha_{pl}} \times {\rm ff}
\end{equation}
where ff is the filling factor of plages, the subscript "pl" corresponds to the local emission Ca$_{\rm loc}$ or H$\alpha_{\rm loc}$ in plages according to Eq.~\ref{eqcram}, the subscript "qs" correspond to the local emission Ca$_{\rm loc}$ or H$\alpha_{\rm loc}$ outside plages (quiet star) according to the same equation, and subscript 0 refers to a constant basal flux over time. Ca$_{\rm qs}$ is in principle very small. Ca$_0$ and H$\alpha_0$ should take values in the proximity of the upper left points in the H$\alpha$ versus Ca plots in Fig.~\ref{indmoyhaca} (i.e. close to the most quiet stars in the sample, for a given T$_{\rm eff}$ bin). The temporal dependence is not explicitly written here for simplification, and is specified in the following sections. 
A noise typical of the observed time series is added to these synthetic indices. Different tests are then conducted in the following sections according to those models.

\begin{figure*}
\includegraphics{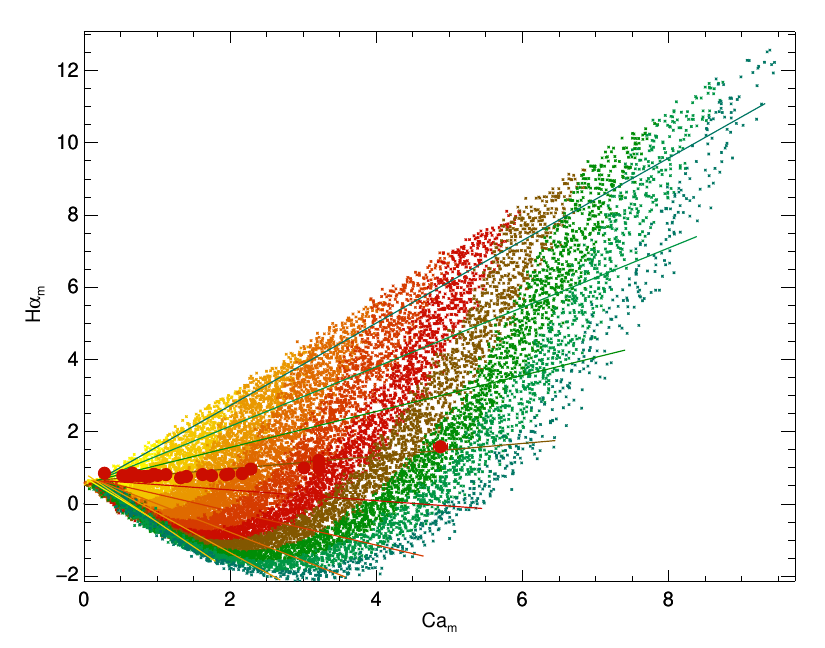}
\caption{Example of the Ca$_{\rm m}$ and H$\alpha_{\rm m}$ emission covered by the simulation (Test \#1) for a given ($x_0$,$y_0$) pair ($x_0$=5,$y_0$=-5). The colour code corresponds to increasing average ff (from 0.001, yellow, to 0.6, blue). Only a fraction of the points are shown for practical reasons (the original grid is 10 times denser). Superposed straight lines indicate the position of constant Ca$_{\rm pl}$ values (from 1.1 in yellow to 15.6 in blue). The observed points for stars in the 3600-3700~K bins are represented in red. 
}
\label{exsimu_ca_ha}
\end{figure*}

\subsection{Test \#1: Temporal variability based on ff(t) only}
\label{sec53}

In this first test, we consider that for a given star, the variability in Eqs.~\ref{eqtest1_ca} and \ref{eqtest1_ha} is due to  ff alone, while Ca$_{\rm pl}$ is always the same for that star (i.e. all plages of a given star have the same Ca emission per unit surface), and Ca$_{\rm qs}$ is the same for all stars. We consider a generic ff(t), varying between a minimum and maximum values, where the variability could be due to rotation or structure evolution.

We first consider that Ca$_{\rm qs}$=0. In this case, we can easily write ff(t) as a function of $S_{\rm Ca}(t)$ from Eq.~\ref{eqtest1_ca}, which, combined with Eq.~\ref{eqtest1_ha}, provides an expression for $S_{\rm H\alpha}(t)$ versus $S_{\rm Ca}(t)$. The slope is equal to ${\rm H\alpha_{pl}/ Ca_{pl}}$
and therefore has the same sign as H$\alpha_{\rm pl}$. Different regimes, corresponding to different Ca$_{\rm pl}$ values and ff ranges, are illustrated in the lower panel in Fig.~\ref{ca_ha_schema}. All stars with Ca$_{\rm pl}>2x_0$ have positive slopes, and all stars with Ca$_{\rm pl}<2x_0$ have negative slopes. The global pattern is therefore incompatible with observations: in order to obtain both positive and negative slopes in the quiet star regime, we need to have stars with a small ff(t) and a large Ca$_{\rm pl}$ above the H$\alpha_0$ level, which is not what we observe, since stars with different slope signs are below that level. 
The offset is equal to ${\rm H\alpha_0 -  Ca_0 H\alpha_{pl}/ Ca_{pl}}$, and is a decreasing function of ${\rm Ca_{pl}}$, which is compatible with observations.

We now relax the Ca$_{\rm qs}$=0 assumption. If Ca$_{\rm qs}$ is different from zero and the same for all stars, the different straight lines (which are shown in the lower panel of Fig.~\ref{ca_ha_schema}) are shifted downwards. Lines corresponding to different values of Ca$_{\rm pl}$ will not cross each other in exactly the same point due to the (1-ff) factor however. If Ca$_{\rm qs}$  is small, we expect this effect to be small as well, and this may not be sufficient to shift the high Ca$_{\rm pl}$  lines in the low activity regime sufficiently to reach H$\alpha_{\rm m}$ level below H$\alpha_0$. To do that, we may need a high Ca$_{\rm qs}$ value for those more active stars, which would strongly shift the lines downward. This may be expected if we extrapolate from the solar configuration, for which there is significant flux (chromospheric network and internetwork) outside plages \cite[][]{meunier18a}, which could be higher for more active stars. However, doing so will also shift the points corresponding to those stars to high Ca$_{\rm m}$ as well, and therefore may not explain the diversity of slopes (of both signs) in the low activity regime. 

To better understand the properties of this simple model and estimate its limitations, we computed time series according to Eqs.~\ref{eqtest1_ca} and \ref{eqtest1_ha}. The details are given in App.~\ref{appF1}. We explored a range of ($x_0$,$y_0$) parameters compatible with the range covered by Ca$_{\rm m}$ and H$\alpha_{\rm m}$ level and constraints on ff for the two T$_{\rm eff}$ bin 3500-3600 and 3600-3700 K (chosen in the negative slope regime and with a large number of stars, respectively 38 and 27). For each set of ($x_0$,$y_0$), we explored a range in ff (in practice between minimum and maximum values representing a range covered over time by the emission of a star) and Ca$_{\rm pl}$, and then compare the properties of these time series with observations. 
An example of such a comparison is shown in Fig.~\ref{exsimu_ca_ha}.
We find that these simple models does not allow to reproduce the observed complexity, since correlations are always very close to 1 or -1 only. In addition, individual slopes are not well reproduced either when considering the same ($x_0$,$y_0$) pair for all stars in a given T$_{\rm eff}$ bin. However, other general properties could be reproduced, such as the typical range covered by the slopes, as illustrated in App.~\ref{appF1}. In addition, we concluded that it is  likely that there is an underlying relation between ff and Ca$_{\rm pl}$, larger ff being associated to stronger local emission.

\subsection{Towards a better agreement with observed slopes}

We found that although the statistical behaviour of the slopes (H$\alpha$ vs Ca) was in reasonable agreement with observations (proper dispersion), they never matched exactly each observation, leading to wrong offsets (defined as the H$\alpha$ value for a Ca index of 0 when performing a linear fit.) as well. 
A first improvement of the simple model used in Test \#1 would be to consider an additional source of temporal variability: in addition to ff(t), plages properties, namely Ca$_{\rm pl}$(t), can be different over time for a given star (i.e. all plages do not have the same properties). In this case, the expected variability over time is therefore no more the straight line  illustrated in Fig.~\ref{ca_ha_schema} for Test \#1, but is non-linear. This would therefore affect the slopes. The precise impact however depends on the exact relationship chosen between Ca$_{\rm pl}$(t) and ff(t).

Another possibility to improve the agreement with the observed slope would be to consider that Ca$_{\rm qs}$ depends on the star. In Test \#1, we considered that stars in a given T$_{\rm eff}$ bin share the same properties in terms of $x_0$ and $y_0$ (same law from Eq.~\ref{eqcram}) but also the same amount of emission in the quiet star chromosphere: they only differed by the plage properties (different Ca$_{\rm pl}$ regimes). Even if it was small, the quiet star emission already played an important role on the slope. If we relax the condition that Ca$_{\rm qs}$ is the same for all stars, this would allow to shift the straight line in Fig.~\ref{ca_ha_schema} to a different level for each star, which could increase the mixing in slope sign for a given Ca$_{\rm m}$ and allow for an adjustment for each star. This would for example be expected from the solar case compared to other solar-type stars, since the emission in the quiet chromosphere must change from one star to the other to explain the observations \cite[][]{meunier18a}. A comparison between the retrieved slopes and the observed one in Test \#1 shows that the difference between the two is significantly lower, and in general small, compared to |$y_0$|, showing that changing slightly Ca$_{\rm qs}$ might be sufficient to retrieve the proper slopes.

\subsubsection{Test \#2: towards a better agreement with observed correlations}

The results obtained in Test \#1 also led to the important conclusion that the simple models based on two components can not reproduce the observed correlations different from 1 or -1. 
An additional level of complexity for a given star then appears to be necessary to reproduce the correlation which we studied in Sect.~4. 
In this section, we therefore check if a complex activity pattern consisting of different plages with different properties for a given star, still associated to the law in Eq.~\ref{eqcram}, could help reconcile observations and models, especially for the correlations.
If plages of a given star have different values of Ca$_{\rm pl}$ and H$\alpha_{\rm pl}$, they may correspond to different regimes law described in Eq.~\ref{eqcram}: Some of those plages then could be in the absorption regime, while others are at the same time in the emission regime, leading to some complex temporal variability and a departure from a good correlation between the Ca and H$\alpha$ time series. 

It is beyond the scope of this paper to cover all possible parameters, so we chose to explore the possible space with a few values. We chose an average total filling factor (over the sphere) of four different levels (2\%, 5\%, 10\%, and 30\%) and considered three ranges for plage sizes. For each of these 12 configurations, we performed 1000 realisations of the time series for each set of parameters, in which ($x_0$,$y_0$) and $\alpha$ (the slope between  Ca$_{\rm pl}$  and ff over the stars in the T$_{\rm eff}$ bin) were varied randomly, in the range used for Test \#1. We also assumed that Ca$_{\rm qs}$=0 for simplification. 
Each  time series is built as follows. We use a time step, over of 1 day, over 1000 days, and a rotation period of 25 days. The star is seen edge-on. Structures of various sizes and lifetimes are added over time depending on how many previous plages have been eliminated, to keep close to the objective in terms of total ff. 
Ca$_{\rm pl}$ depends on the ff of each plage as in equation 6. H$\alpha_{\rm pl}$ can therefore be positive or negative depending on its size and $\alpha$. There are therefore usually plages in both regimes at the surface of the star, in proportion that depends on the parameters (mostly $\alpha$ and range in size). The range of parameters considered here corresponds to configurations with a number of plages at a given time between typically 4 and 40. 

Qualitatively, we find that low $\alpha$ and ff of individual plages, as well as the total ff,  control the resulting Ca$_{\rm m}$. Conversely to simulations performed in the previous sections, it is possible to reach any correlation, including close to zero, thanks to the presence of plages in different regimes at the same time. Very low values of $\alpha$ always lead to negative correlation, between 0 (very low $\alpha$) and -1 (moderate $\alpha$). Above a certain threshold, which depends on the other parameters, positive correlations can be obtained as well as negative ones. Finally, in the most active states, only positive correlations are reached, which is compatible with observations. However, for these few configurations, although there are domains of Ca$_{\rm m}$ where both signs of the slope coexist, as observed, this never occurs for the lowest activity level (typically below 1): a larger range of parameters should be explored to see if this could be produced in the activity range corresponding to observations.

\section{Conclusion}

We studied the relation between three chromospheric indices from a large sample of M stars   in detail; namely Ca II H \& K, Na D1 and D2, and H$\alpha$, first from their time-averaged values and then in more detail from the relationship between the corresponding time series. The originality of our approach lies in the fact that we consider the relationship on short and long timescales in order to find clues about the differences between their behaviour. In addition, we compared the observed properties with our findings for FGK stars in \cite{meunier22}, and with simple models. Our main results can be summarised as follows: 

\begin{itemize}
\item{Our analysis of the averaged indices reveals a U-shaped lower envelope for the H$\alpha$ versus Ca and  H$\alpha$  versus  Na relationships, and a significantly negative slope in the lowest-activity regime. This may be compatible with expectations based on the findings of  \cite{cram79}. However, we observe this behaviour only for M dwarfs with T$_{\rm eff}$ above 3400-3500~K; that is, in the partially convective regime. The quietest stars in our sample also always exhibit significant emission in the core of the Ca II H \& K lines, meaning that no star is completely inactive. Finally, the Na versus Ca relationship exhibits a different slope in the two regimes (quiet and active stars), again for partially convective stars, which may not be related to the variability of the star but rather to its fundamental parameters. }
\item{The correlations between the three activity indices are very diverse. These correlations are not significantly impacted by the choice of H$\alpha$ bandwidth, as opposed to the case of solar-type stars \cite[][]{gomes22}. We find stars for which all indices are well correlated with each other: for 44\% of the stars, both correlations are above 0.5, and they are above 0.8 for 13\% of the stars. There are also stars for which only the Ca and H$\alpha$ indices are correlated, or only the Ca and Na indices, with many different configurations when considering the three pairs of indices, highlighting the complex behaviour between the three indices. The degradation of the Ca--H$\alpha$ and Ca--Na correlations does not correspond to stars with similar activity levels, and may therefore involve different processes.
Concerning the relation between Ca and H$\alpha$, we find only a few anticorrelated stars (below 5\% of the sample); this is the case for FGK stars, but compared to these latter, many fewer of the stars studied here show a correlation close to zero. However, similarly to the case for FGK stars \cite[][]{meunier22}, we find no dependence of these correlations on spectral type; the low correlations or anticorrelations tend to occur for relatively quiet stars (very low $\log R'_{HK}$) but 
still with significant variability. The dependence on metallicity is complex, because we show that the relation between metallicity and activity level depends on T$_{\rm eff}$ (this may be due to a slower breaking for later-type stars, leading to a different proportion of active stars at a given age), as opposed to the case of FGK stars, pointing to different processes: metallicity and $\log R'_{HK}$ are more strongly correlated for quiet stars and in the T$_{\rm eff}$  regime where there is a negative slope in the average index relationship.
The star with the strongest anticorrelation exhibits much higher H$\alpha$ variability compared to the Ca variability than in all other stars.}
\item{As opposed to the case for FGK stars, there is little difference in the correlations and variability between the short and long timescales. This suggests that the variability could in principle be more easily explained by the presence of plages without adding another process \cite[such as the filaments proposed for FGK stars;][]{meunier09a,meunier22}. The decrease in H$\alpha_{\rm m}$ versus Ca$_{\rm m}$ in the low-activity regime could be related to the theoretical work of \cite{cram79}. However, some stars in that regime have a positive slope (variability in H$\alpha$ versus Ca). In addition, for a given star, the correlation computed for different seasons may vary greatly, suggesting that the star may be in different regimes over time.   }    
\item{Simple toy models show that the link between such a local law and the integrated indices over the whole disc is complex, especially when considering all observables. We observe that stars in a regime where the H$\alpha_{\rm m}$ does not  change significantly with activity level are in fact variable in time in H$\alpha$. The interpretation in terms of a stronger absorption in H$\alpha$ when activity increases before going to emission at a higher activity level would necessitate the presence of complex activity patterns on the surface of some of those stars to explain such an observation. It also suggests that plages with different properties exist for a given star. There also appears to be an underlying relationship between the local emission and the filling factor of plages. Basal fluxes (corresponding to no magnetic activity) that are  different from one star to the other may also be necessary to explain the observations.  }    
\end{itemize}

We will therefore implement more complex models in the future to better explain the relation between Ca and H$\alpha$ variability. This should also allow us to introduce long-term variability and to compare the short- and long-timescale variability and correlations.

Finally, for about half of the stars, all three indices are not well correlated. Therefore, when using one particular index to eliminate false positives when searching for exoplanets in radial velocity, one should be cautious because a lack of activity variability at a certain period in a given index does not guarantee a lack of activity variability in  the other indices:  one should therefore always consider a larger panel of indices when studying relationships with radial velocity.

\begin{acknowledgements}

This work has been supported by a grant from LabEx OSUG@2020 (Investissements d'avenir - ANR10LABX56).
This work was supported by the "Programme National de Physique Stellaire" (PNPS) of CNRS/INSU co-funded by CEA and CNES.
This work was supported by the Programme National de Planétologie (PNP) of CNRS/INSU, co-funded by CNES.  
The HARPS data have been retrieved from the ESO archive at http://archive.eso.org/wdb/wdb/adp/phase3\_spectral/form.
This research has made use of the SIMBAD database, operated at CDS, Strasbourg, France. 
ESO program IDs (the name of the PI is indicated) corresponding to data used in this paper are: 
 095.C-0718 (Albrecht)
 096.C-0082 191.C-0505 (Anglada-Escude) 
 100.C-0884 (Astudillo Defru) 
 0101.D-0494 (Berdinas) 
 082.C-0718 180.C-0886  183.C-0437 191.C-0873 198.C-0873 1102.C-0339 (Bonfils) 
 078.D-0245 (Dall) 
 075.D-0614 (Debernardi)
 096.C-0499  098.C-0518 0100.C-0487 (Diaz) 
 076.C-0279 (Galland) 
 074.C-0037     075.C-0202 076.C-0010 (Guenther) 
 096.C-0876 097.C-0390 (Haswell) 
 078.C-0044     (Hébrard)
 097.C-0090 (Kuerster)   
 0104.C-0863 (Jeffers)
 089.C-0006 (Lachaume) 
 089.C-0739 098.C-0739 099.C-0205 0104.C-0418 192.C-0224 (Lagrange)  
 097.C-0864 (Lannier) 
 085.C-0019 087.C-0831 089.C-0732 090.C-0421 091.C-0034 093.C-0409 095.C-0551 096.C-0460 098.C-0366 099.C-0458 0100.C-0097 0101.C-0379  0102.C-0558 0103.C-0432  196.C-1006 (LoCurto)
 072.C-0488  077.C-0364 (Mayor) 
 076.C-0155 (Melo)  
 089.C-0050 (Pepe) 
 185.D-0056 (Poretti)  
 074.C-0364 (Robichon) 
 084.C-0228 090.C-0395 (Ruiz)    
 086.C-0284 (Santos) 
 079.C-0463 (Sterzik) 
 0100.C-0414 (Trifonov) 
 183.C-0972 192.C-0852 (Udry) 
 060.A-9036 60.A-9709.
\end{acknowledgements}

\bibliographystyle{aa}
\bibliography{biblio}

\begin{appendix}


\section{Sample and correlations}
\label{appA}

\begin{table}[h]
\caption{Stellar sample}
\label{tab_targets}
\begin{center}
\renewcommand{\footnoterule}{}  
\begin{tabular}{lllllllllllll}
\hline
Name  &  V-K & T$_{\rm eff}$ & FeH & Number & $\log R'_{HK}$ & $S_{\rm Ca}$ & rms & $S_{\rm Na}$ & rms & $S_{H\alpha}$ & rms & Season \\
          &       & (K)  &  & of nights &  &  & $S_{\rm Ca}$ &  & $S_{\rm Na}$ &  & $S_{H\alpha}$ &  \\
 \hline
CD-246144 & 3.55 & 3922 & -0.17 &   12 &  -4.62 &   1.18 &   0.10 &   0.13 &   0.00 &   0.66 &   0.01 &    \\
CD-4114656 & 3.41 & 3925 & -0.68 &   18 &  -4.67 &   0.97 &   0.05 &   0.14 &   0.00 &   0.67 &   0.01 &    \\
GJ1 & 4.03 & 3589 & -0.45 &   38 &  -5.51 &   0.41 &   0.05 &   0.10 &   0.00 &   0.82 &   0.01 &    \\
GJ43 & 6.31 & 3616 &  - &   17 &  -6.38 &   0.65 &   0.18 &   0.13 &   0.01 &   0.79 &   0.01 &    \\
GJ54.1 & 5.65 & 3200 & -0.40 &  202 &  -4.59 &   7.07 &   2.38 &   0.48 &   0.12 &   1.84 &   0.37 &  (S) \\
GJ87 & 3.96 & 3700 & -0.31 &  136 &  -5.36 &   0.49 &   0.09 &   0.13 &   0.00 &   0.81 &   0.01 &  (S) \\
GJ91 & 4.23 & 3757 &  0.01 &   21 &  -4.89 &   1.17 &   0.10 &   0.17 &   0.00 &   0.76 &   0.01 &    \\
GJ93 & 3.78 & 3860 &  - &   12 &  -5.17 &   0.61 &   0.06 &   0.15 &   0.00 &   0.75 &   0.01 &    \\
GJ105B & 5.03(d) & 3200 & -0.02 &   20 &  -5.51 &   0.75 &   0.11 &   0.10 &   0.01 &   0.89 &   0.02 &    \\
GJ126 & 3.87 & 3830 & -0.29 &   29 &  -5.00 &   0.76 &   0.09 &   0.15 &   0.01 &   0.76 &   0.01 &    \\
GJ149B & 2.66 & 4149(*) & -0.16 &   29 &  -4.69 &   0.54 &   0.08 &   0.11 &   0.00 &   0.54 &   0.01 &    \\
GJ157B & 4.55 & 3545 & -0.02 &   10 &  -4.44 &   6.83 &   0.91 &   0.37 &   0.02 &   2.09 &   0.08 &    \\
\hline
\end{tabular}
\end{center}
\tablefoot{
Main parameters of the stars in our  sample and activity from our analysis. 
V-K values are from the CDS when available, or derived from the relationship between G-K versus V-K when V is not available (indicated by "d", concerns 15 stars), as in \cite{mignon21c}. 
T$_{\rm eff}$ values are from the CDS. When missing, the reported value, indicated by (*), means that we have estimated it from a T$_{\rm eff}$ versus V-K linear law (see text, concerns 10 stars). The metallicities (FeH) are from \cite{casagrande08}, \cite{neves13},  \cite{kordopatis13}, \cite{gaidos14}, \cite{gaspar16}, \cite{houdebine16}, \cite{aganze16}, \cite{passegger18},  \cite{maldonado19}, \cite{hojjatpanah19},  \cite{kuznetsov19}, \cite{maldonado20}, \cite{birky20}, \cite{steinmetz20}, \cite{hojjatpanah20},  \cite{jonsson20}, \cite{sarmento21}, \cite{buder21},  \cite{marfil21}, and \cite{HubbardJames22}.   
The number of nights, average $\log R'_{HK}$, average of the three indices and their rms are from the present paper. The flag (S) indicates if the star is in the subsample with more than four seasons (Sect.~\ref{sec24}). Only the beginning of the table is shown here, the full table is available at the CDS.
}
\end{table}

\begin{table}[h]
\caption{Global correlations and slopes}
\label{tab_correl}
\begin{center}
\renewcommand{\footnoterule}{}  
\begin{tabular}{lllllll}
\hline
Name  &  C(Ca-H$\alpha$) & C(Na-Ca)  & C(Na-H$\alpha$) & Slope H$\alpha$ versus Ca  & Slope Na versus Ca  & Slope H$\alpha$ versus Na \\ 
 \hline
 CD-246144 &  0.903$\pm$0.028 &  0.839$\pm$0.056 &  0.772$\pm$0.056 &  0.085$\pm$ 0.006 &  0.024$\pm$ 0.003 &  2.550$\pm$ 0.290 \\
CD-4114656 &  0.259$\pm$0.130 &  0.497$\pm$0.145 &  0.343$\pm$0.102 &  0.033$\pm$ 0.018 &  0.023$\pm$ 0.007 &  1.016$\pm$ 0.300 \\
GJ1 & -0.156$\pm$0.015 &  0.900$\pm$0.010 & -0.278$\pm$0.017 & -0.056$\pm$ 0.003 &  0.068$\pm$ 0.001 & -1.035$\pm$ 0.053 \\
GJ43 &  0.061$\pm$0.172 &  0.072$\pm$0.217 & -0.715$\pm$0.111 & -0.000$\pm$ 0.006 &  0.003$\pm$ 0.005 & -0.926$\pm$ 0.154 \\
GJ54.1 &  0.868$\pm$0.003 &  0.858$\pm$0.004 &  0.938$\pm$0.001 &  0.135$\pm$ 0.001 &  0.044$\pm$ 0.000 &  2.880$\pm$ 0.009 \\
GJ87 & -0.327$\pm$0.015 &  0.685$\pm$0.016 & -0.303$\pm$0.015 & -0.050$\pm$ 0.002 &  0.043$\pm$ 0.001 & -0.859$\pm$ 0.042 \\
GJ91 &  0.663$\pm$0.041 &  0.687$\pm$0.052 &  0.562$\pm$0.046 &  0.067$\pm$ 0.004 &  0.030$\pm$ 0.002 &  1.405$\pm$ 0.117 \\
GJ93 & -0.052$\pm$0.145 &  0.212$\pm$0.174 &  0.743$\pm$0.101 & -0.013$\pm$ 0.016 &  0.012$\pm$ 0.009 &  1.608$\pm$ 0.302 \\
GJ105B &  0.671$\pm$0.070 &  0.200$\pm$0.130 &  0.195$\pm$0.060 &  0.109$\pm$ 0.014 &  0.009$\pm$ 0.006 &  0.334$\pm$ 0.130 \\
GJ126 &  0.016$\pm$0.109 &  0.676$\pm$0.089 & -0.030$\pm$0.091 & -0.013$\pm$ 0.008 &  0.035$\pm$ 0.005 & -0.228$\pm$ 0.125 \\
GJ149B &  0.542$\pm$0.046 &  0.849$\pm$0.037 &  0.594$\pm$0.057 &  0.043$\pm$ 0.003 &  0.029$\pm$ 0.002 &  1.377$\pm$ 0.125 \\
GJ157B &  0.467$\pm$0.026 &  0.598$\pm$0.037 &  0.847$\pm$0.019 &  0.055$\pm$ 0.003 &  0.021$\pm$ 0.001 &  2.815$\pm$ 0.099 \\
\hline
\end{tabular}
\end{center}
\tablefoot{
Global correlations and slopes for the 177 stars in our sample, with their 1-$\sigma$ uncertainty. Only the beginning of the table is shown here, the full table is available at the CDS. 
}
\end{table}

\section{Examples of spectra}
\label{appB0}

\begin{figure*}[h!]
\includegraphics{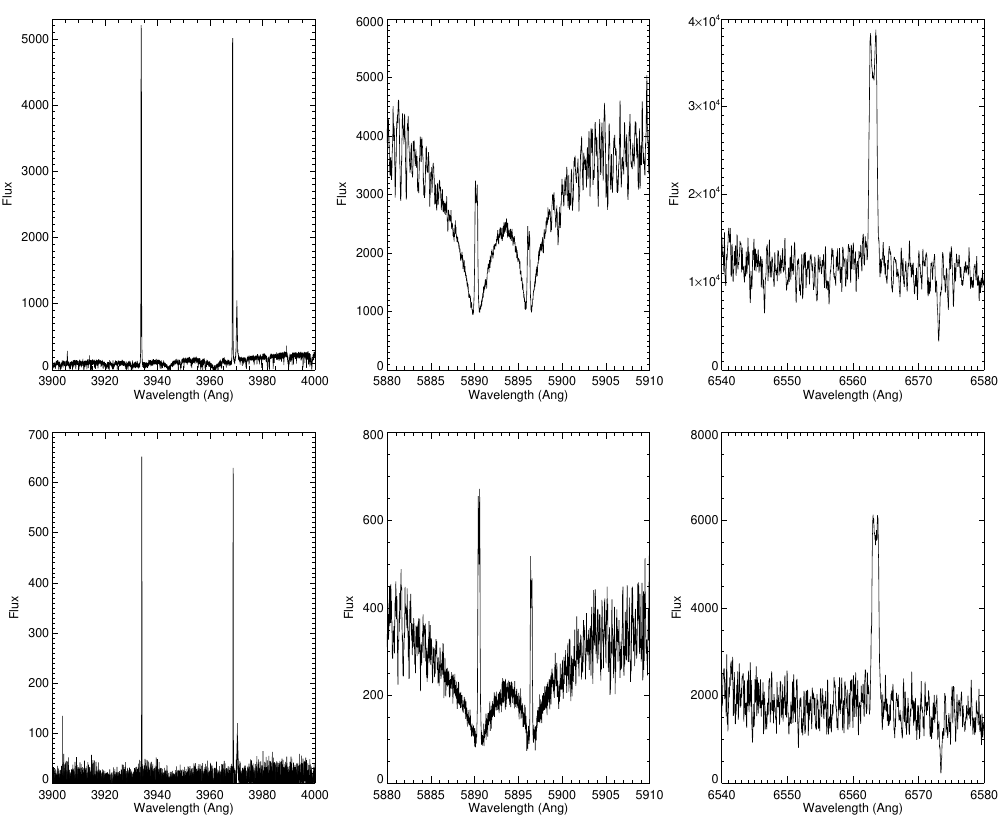}
\caption{
Example of spectra in the three bands: Ca II H \& K (left panels), Na  D1 and D2 (middle panels), and H$\alpha$ (left panels), corresponding to a high S/N (GJ388, upper panel, S/N around 10 in the Ca band) and a low S/N (GJ54.1, lower panels, S/N around 1 in the Ca band).
}
\label{exsnr}
\end{figure*}


\section{Relationship between average indices}
\label{appB}

\begin{figure*}[h!]
\includegraphics{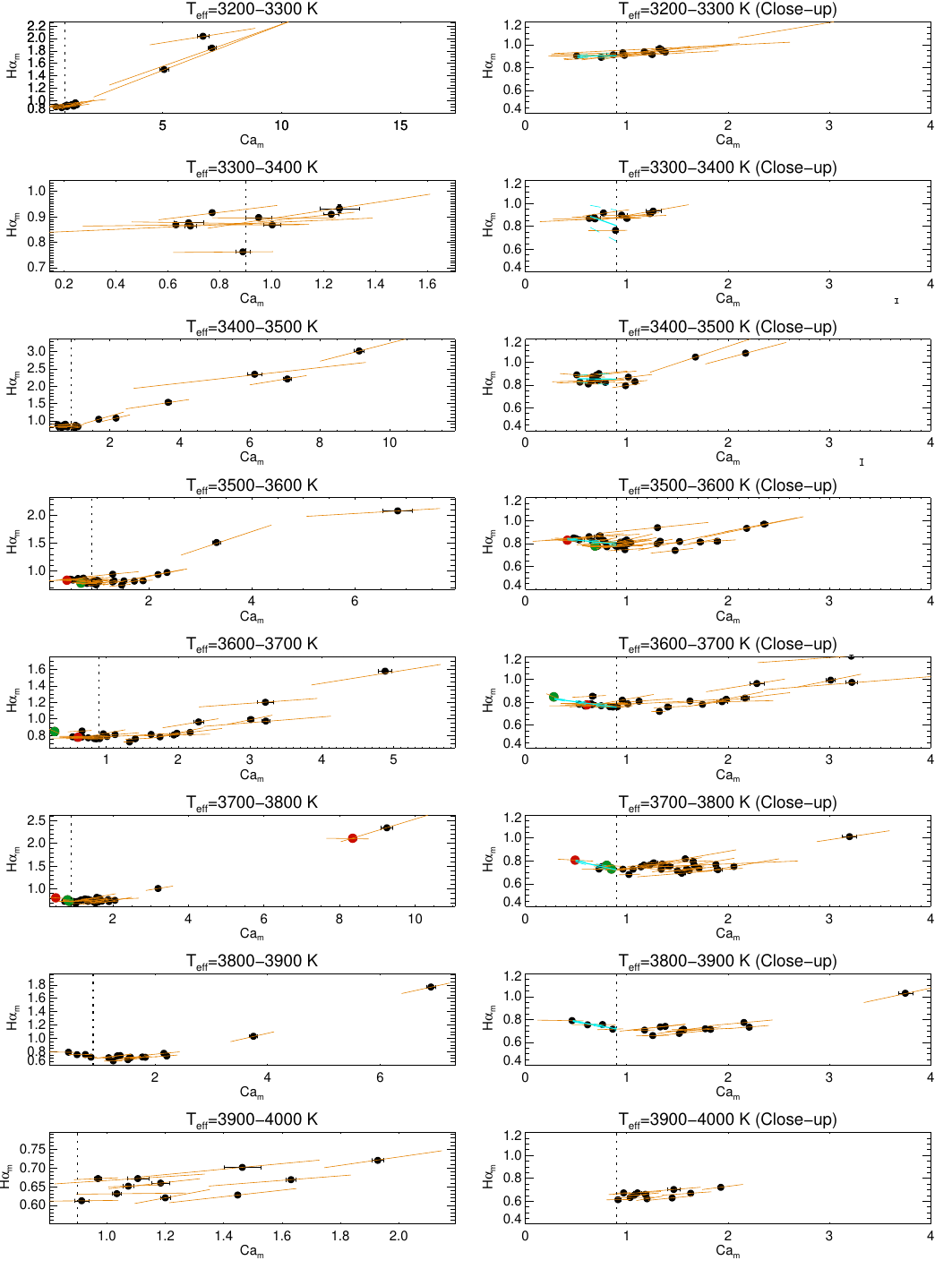}
\caption{
Relation between H$\alpha_{\rm m}$ and Ca$_{\rm m}$  in different T$_{\rm eff}$ bins (from top to bottom), for all stars (left panels) and without the most active stars (right panels). Circles in green and red correspond to stars with anticorrelation between indices (see Sect.~\ref{sec4}, below -0.5 and -0.3 respectively). The orange straight lines correspond to the linear fit between indices for that star, plotted for the covered range in $S_{\rm Ca}$(t).}
\label{indmoyhaca}
\end{figure*}

\begin{figure*}[h!]
\includegraphics{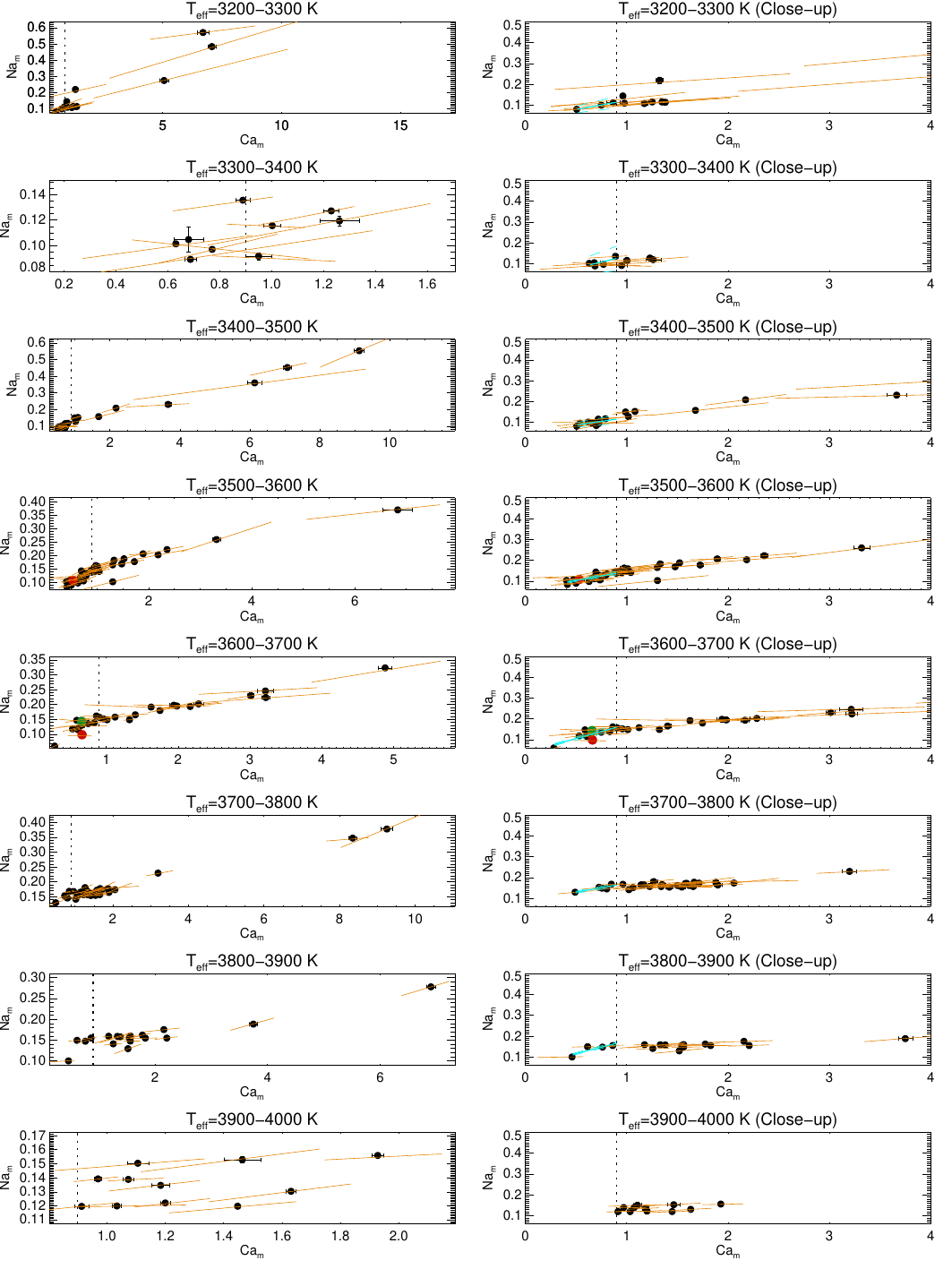}
\caption{Same as Fig.~\ref{indmoyhaca} for Na$_{\rm m}$ versus Ca$_{\rm m}$.}
\label{indmoynaca}
\end{figure*}

\begin{figure*}[h!]
\includegraphics{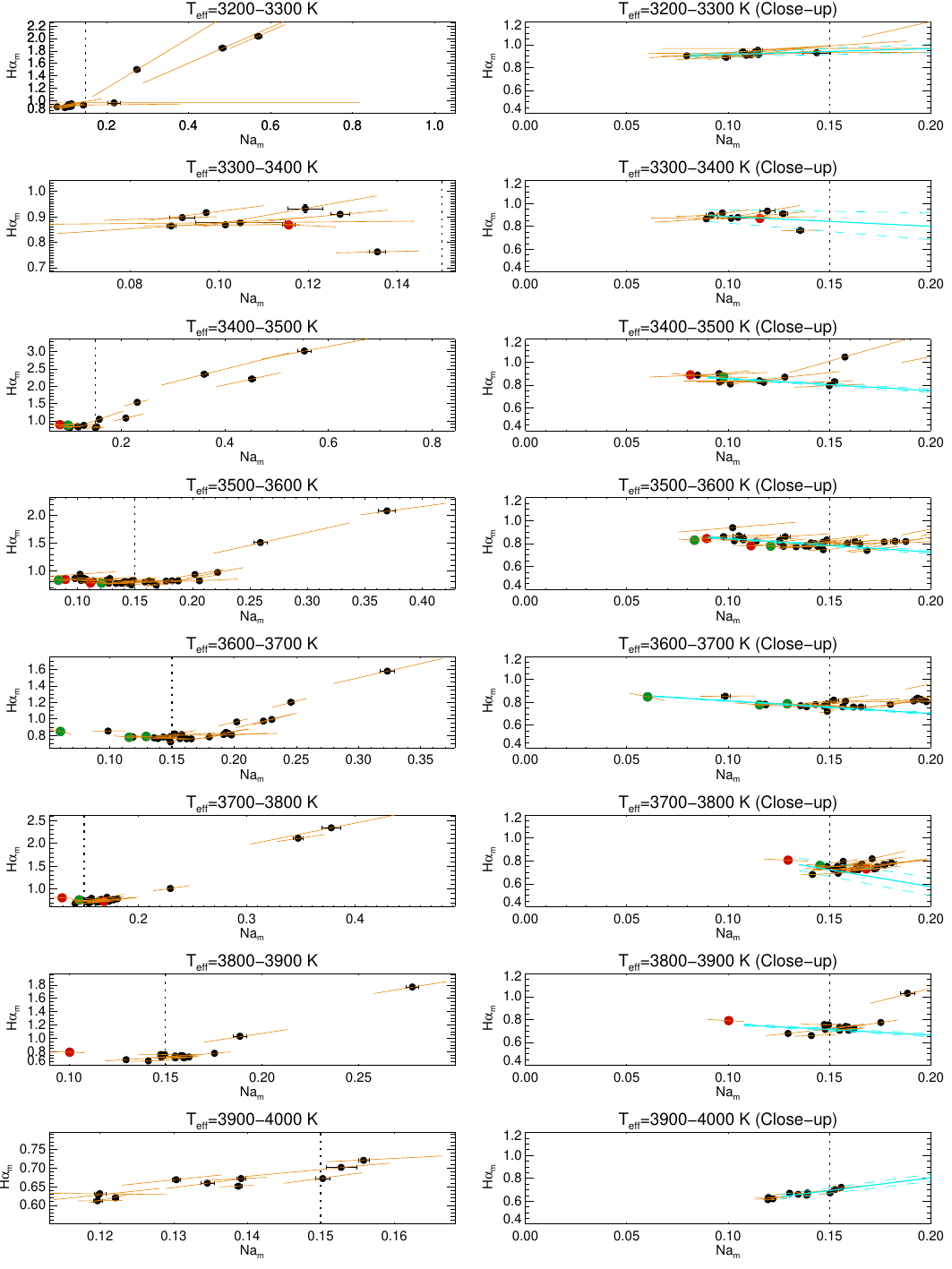}
\caption{Same as Fig.~\ref{indmoyhaca} for H$\alpha_{\rm m}$ versus Na$_{\rm m}$.}
\label{indmoyhana}
\end{figure*}

Figures~\ref{indmoyhaca} (H$\alpha_{\rm m}$ versus Ca$_{\rm m}$), \ref{indmoynaca} (Na$_{\rm m}$ versus Ca$_{\rm m}$), and \ref{indmoyhana} (H$\alpha_{\rm m}$ versus Na$_{\rm m}$) show the relation between the three pairs of averaged indices for all T$_{\rm eff}$ bins. They are studied in Sect.~\ref{sec3}.

\newpage

\section{Examples of time series}
\label{appC}

\begin{figure*}
\includegraphics{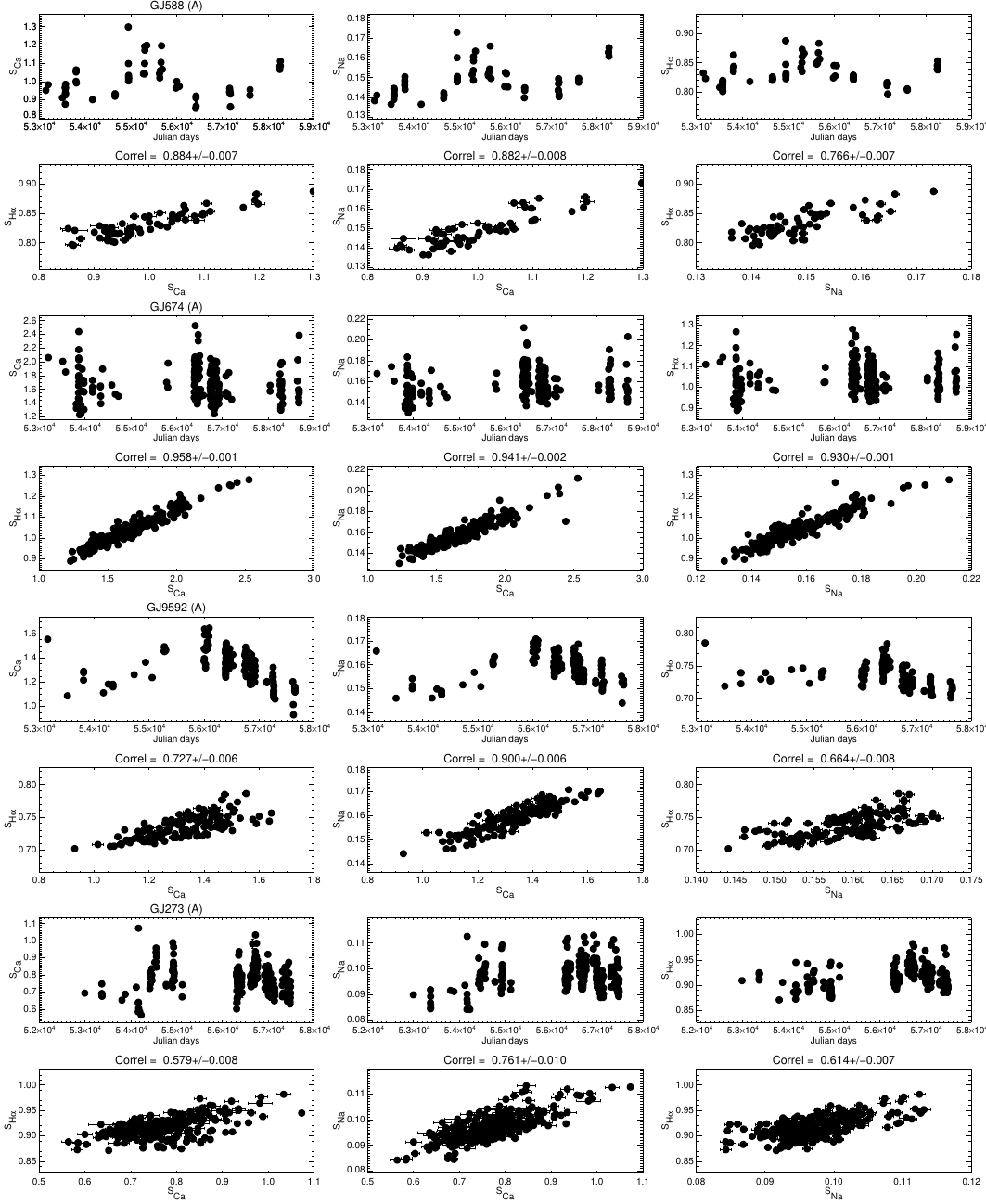}
\caption{Examples of very well (first 2)  or moderately well (last 2) correlated time series (category A).}
\label{exemple1}
\end{figure*}

\begin{figure*}
\includegraphics{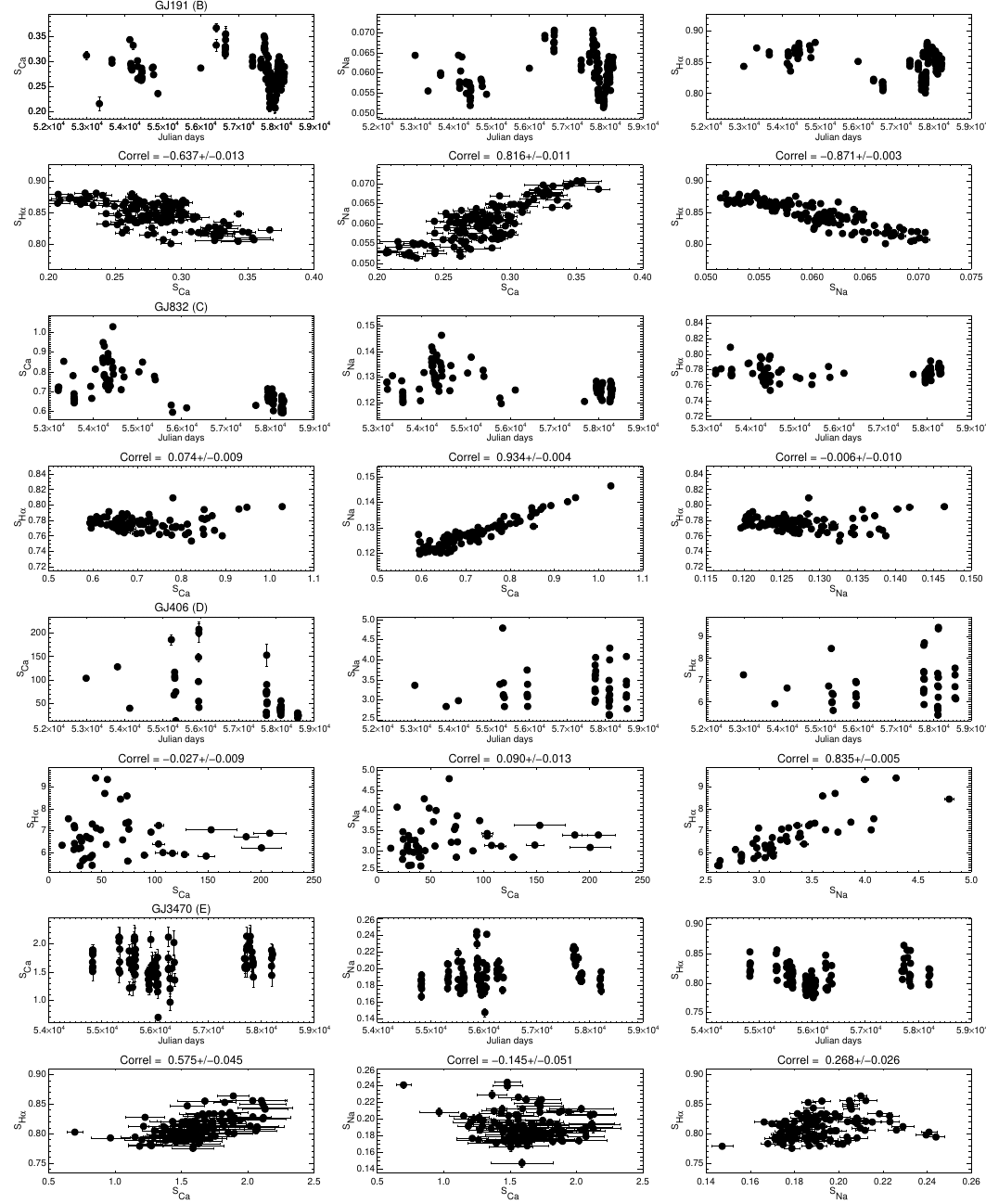}
\caption{Examples of uncorrelated time series (categories B, C, D, and E).}
\label{exemple2}
\end{figure*}

We illustrate the different categories of relationships with a few examples in Fig.~\ref{exemple1} (examples with a good correlation) and Fig.~\ref{exemple2} (examples with a degraded correlation). The  categories (A-E) are represented in the lower panel of Fig.~\ref{relatcorrel} and in the figures in Sect.~\ref{sec3}.

\section{Metallicity effects}
\label{appD}

\begin{figure}
\includegraphics{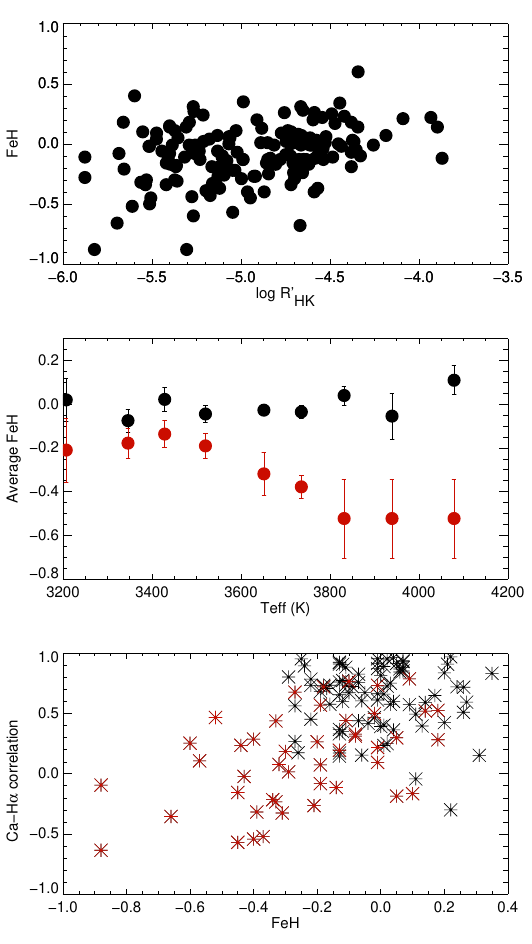}
\caption{FeH versus $\log R'_{HK}$ (upper panel), versus T$_{\rm eff}$ for quiet stars, in red, and active stars, in back (middle panel), and Ca-H$\alpha$ correlation versus FeH separately for quiet and active stars (lower panel).
}
\label{feh}
\end{figure}

\begin{table}
\caption{Relationship with metallicities}
\label{tab_feh}
\begin{center}
\renewcommand{\footnoterule}{}  
\begin{tabular}{lllllll}
\hline
Selection & \multicolumn{2}{c}{FeH$<$-0.2} & \multicolumn{2}{c}{|FeH|$<$0.2} & \multicolumn{2}{c}{FeH$>$0.2}  \\ \hline
           &  M & FGK & M & FGK & M & FGK \\ \hline
Quiet & 22  & 74  & 23  & 160   & 0 & 33 \\
Active & 9 &  9 & 64 &  57  & 9 &  3 \\
\hline
\end{tabular}
\end{center}
\tablefoot{Number of stars with low, medium, and high metallicity in our 3400-3900 K range, for quiet and active stars (127 stars). Similar numbers are computed for FGK stars from Paper II (360 stars for T$_{\rm eff}$ between 5100 and 6100 K).
}
\end{table}

Figure~\ref{feh} shows the metallicity (FeH) versus $\log R'_{HK}$ (weak correlation of 0.32), trend already  observed  by \cite{scandariato17}. There is therefore a trend for the least active stars to be submetallic. 
This is shown more clearly in the middle panel of Fig.~\ref{feh}, showing the average FeH versus T$_{\rm eff}$ separately for quiet and active stars. There is no significant metallicity dependence for the active stars. However, there is a  difference for quiet stars above 3400 K.
Table~\ref{tab_feh} illustrates this for stars in the range 3400-3900K, and compares with FGK, which does not show this behaviour. In fact, interestingly, a plot (not shown here) similar to the middle panel of Fig.~\ref{feh} for FGK stars shows no difference between quiet and active stars, and the correlation between $\log R'_{HK}$ and FeH is -0.15, which is less marked and of the opposite sign compared to M stars. Such a relationship for FGK stars has been shown in \cite{jenkins08} and \cite{meunier17b}, and is weak, and the main difference between active and quiet stars is mostly a larger spread in FeH for quiet stars. 

In addition, a relationship between $\log R'_{HK}$ and FeH could be attributed to very different causes, either at a deep level from different dynamo processes, or affecting the production of spectral lines in the atmosphere. 
First, metallicity could impact the turnover time at the base of the convective zone and therefore differential rotation and the dynamo \cite[][]{bessolaz11,vansaders12,brun17}. Younger stars also tend to be more active and also more metallic. 
On the other hand, metallicity could also affect the contrasts \cite[][]{shapiro14,shapiro15}. \cite{houdebine11} showed that the Ca emission increased with the abondance of Ca II ion, which at first order increases with metallicity.

Coming back to the relationship between our correlations C and FeH, the lower panel of Fig.~\ref{feh} shows the H$\alpha$-Ca correlation as a function of FeH, separately for the quiet stars, with a correlation of 0.46 (a similar computation for FGK stars leads to a negative correlation, with the strongly anticorrelated stars being surmetallic) and the active stars (correlation of -0.04, i.e. no correlation). 
The relationship between activity, correlations with H$\alpha$ and FeH seems to be present for quiet stars only, and is opposite to what is observed for FGK stars. This point toward different dominant processes for M and FGK stars.

\section{Synthetic time series based on Ca observed time series}
\label{appE}

In this  approach, we built H$\alpha$ and Na synthetic time series based on different scalings of the observed Ca time series as in Paper II. Once they are generated, we computed the correlation, slope, and variability as for the observations in the previous section and compared them with the observed ones.

\subsection{Building of the time series}

We explore three assumptions to build the H$\alpha$ and Na synthetic time series. The first assumption (hypothesis A) is based on the properties of stars with perfect correlations between H$\alpha$ (or Na) and Ca. The scaling factor between H$\alpha$ (or Na) and Ca that will allow to build the time series is described below and follows a procedure similar to Paper II. With the second assumption (hypothesis B), the scaling factor between Ca and H$\alpha$ (or Na) time series is the slope computed from the averaged indices in Sect.~\ref{sec42} for each star. Finally, the third assumption (hypothesis C) leads to a scaling  factor equal to the individual slope computed for the star. To build the synthetic time series, the observed Ca time series is  multiplied by the scaling factor corresponding to the chosen hypothesis and index. Some noise is added as prescribed by the observed uncertainty for each star and index, and an offset is added to match the observed level. The details are given in the following sections.

\begin{figure*}
\includegraphics{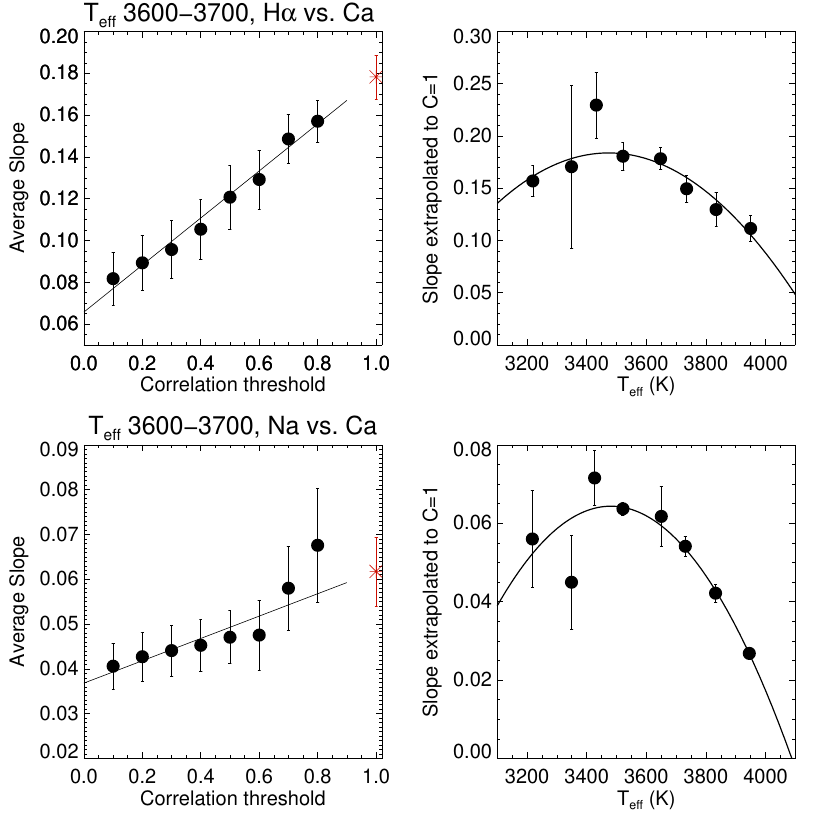}
\caption{Slope versus lower threshold on the correlation for stars in the 3600-3700K range (left panels), with the slope extrapolated to a correlation of 1 in red, and extrapolated slope versus T$_{\rm eff}$ (right panels), in two cases: For the H$\alpha$-Ca relation (upper panels) and for the Na-Ca relation (lower panels). }
\label{loi1}
\end{figure*}

We detail here how hypothesis A is built. We consider here, as in Paper II, the factor relating these emissions for perfectly correlated stars. For this purpose, we compute, for stars in each bin in T$_{\rm eff}$ (100 K), the average slope for different selections of stars, that is with correlation above a certain threshold. An example is given for the 3600-3700 bin in the left panels of Fig.~\ref{loi1}. The resulting relationship is then extrapolated to a correlation of 1, indicated in red. This slope is then used for that T$_{\rm eff}$ bin to convert a Ca emission into a H$\alpha$ emission. They are shown as a function of T$_{\rm eff}$ in the right panels of Fig.~\ref{loi1}.

\subsection{Correlation and slope obtained with synthetic H$\alpha$ and Na time series}

\begin{figure}
\includegraphics{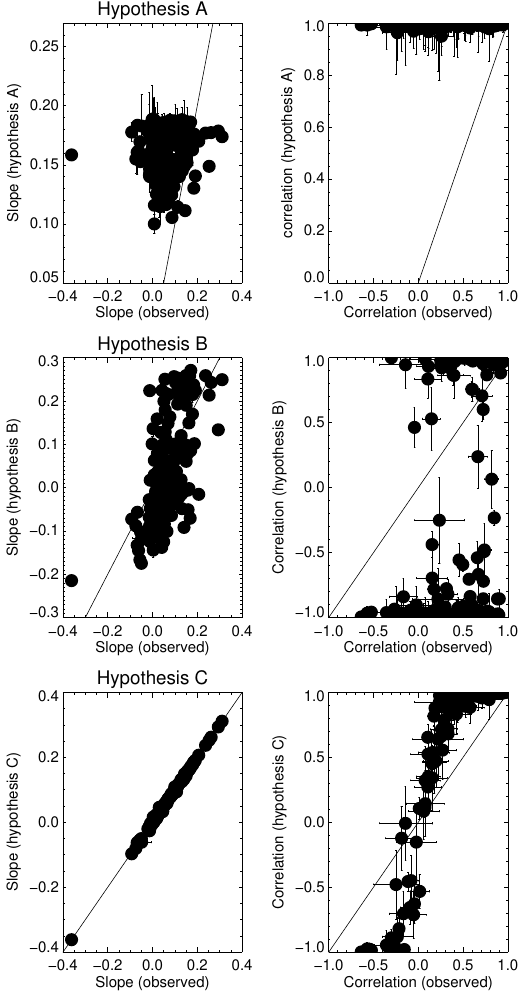}
\caption{Slope versus observed slope (left panels) and correlation versus observed correlation (right panels), for the three hypothesis A, B, and C (from top to bottom). 
}
\label{figcorr_correl_synth}
\end{figure}

With hypothesis A, the correlations are very close to 1, typically higher than 0.9 for Ca-H$\alpha$, and most are above 0.8 (except 10 stars) for Ca-Na. The slopes are on average too high: there is a very large dispersion and no correlation between observed and synthetic slopes.  
Hypothesis B leads to a larger range of correlations,  as shown in Fig.~\ref{figcorr_correl_synth}. However, we observe important differences with observations: 1/ the distribution of the correlations is still strongly biased towards correlations close to 1 for Na-Ca, while it is too strongly biased towards correlations close to -1 for the correlations with H$\alpha$; 2/ for stars with a positive simulated  correlation, the slopes cover a realistic range only for those with correlations very close to 1, otherwise the slopes are much smaller than observed, and there is in fact no strong correlation between the observed and synthetic slopes; 3/ The few stars with a strong synthetic anticorrelation do not show the drop in slope as observed for GJ191. 
Finally, hypothesis C leads to correlations with a slightly better agreement with observations  than with hypothesis B, but they are still always too high (in absolute value). The slopes are naturally in agreement by construction.  

This means that even with a factor typical of the one derived from the time-averaged or individual indices, and taking into account the noise on the data, we lack weak correlations. The relation between indices is therefore more complicated than what is described here in such a simple models, even with a underlying complex variability (from the Ca observations).

\subsection{Expected H$\alpha$ and Na variability from Ca variations}

As in Sect.~\ref{sec434}, we also compute the ST and LT rms for the different indices, and then the ratio between indices at these two timescales, to be compared with the observations in Fig.~\ref{ratiorms}.

For H$\alpha$ versus Ca, hypothesis A leads to a reasonable range for the LT ratio, but the ST ratio does not show a trend with the LT ratio as in the observations. 
Hypothesis C also leads to rms that are often too small. 
On the other hand, the ST versus LT ratio for Ca/Na presents similarities with Fig.~\ref{ratiorms} for the three hypothesis. 

\subsection{Conclusion}

We conclude from these simple simulations that: 1/ they fail to reproduce moderate correlations, even when taking the noise into account, and the factor based on the slopes from averaged indices does not always give the proper sign, and therefore a unique factor applied to the whole time series is  insufficient; 2/ There is not a strong difference between ST and LT behaviours, which is similar to the observations \cite[conversely to the FGK analysis, for which there was a strong difference, which led us to explore the possibility for filaments to explain the observed results][]{meunier09a}, but the amplitude in both H$\alpha$ and Na is often under-estimated.

\section{Toy model protocols for Test \#1}

In this section, we provide details for the toy models discussed in Sect.~\ref{sec53}. 
\label{appF1}

\begin{figure*}
\includegraphics{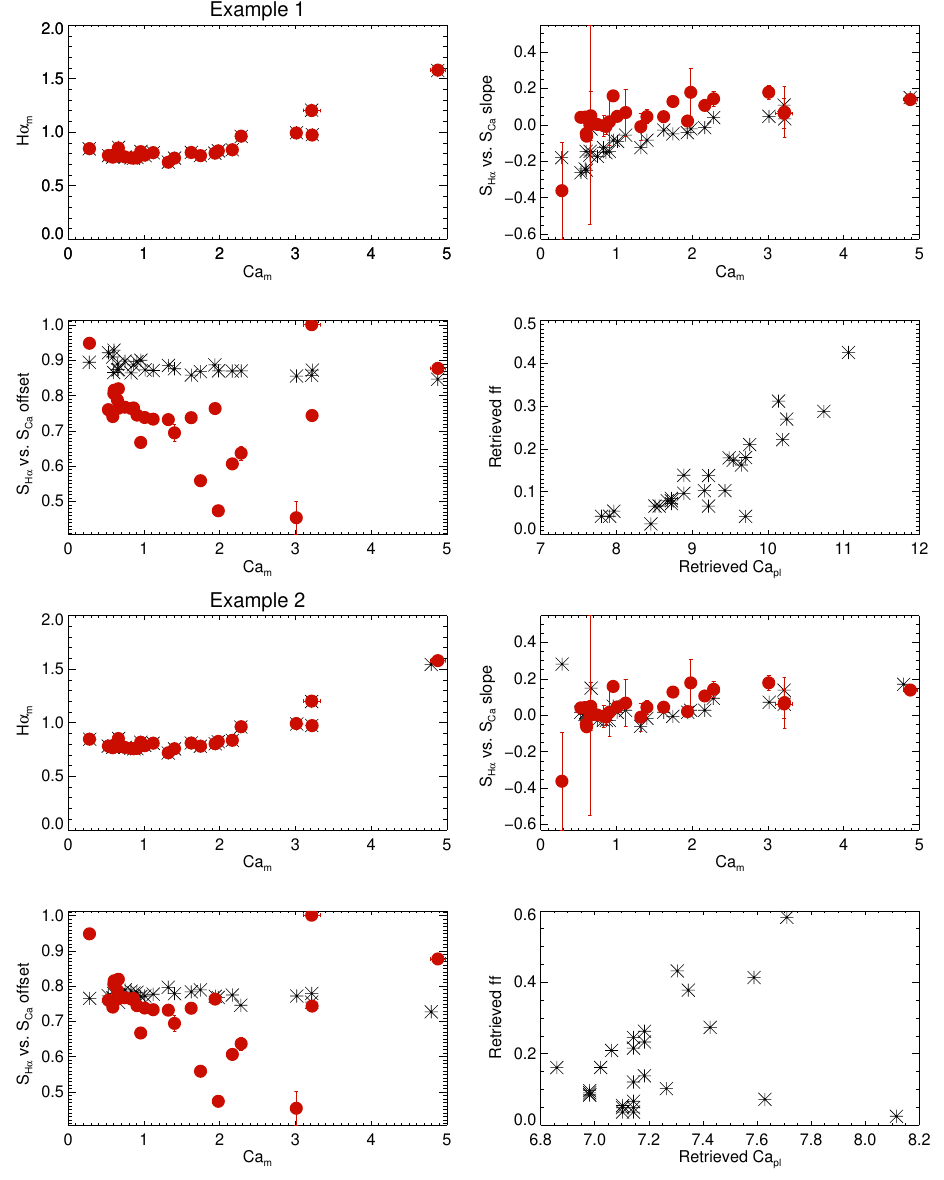}
\caption{Example of a comparison between simulation (Test \#1, selection \#3) and observation: H$\alpha_{\rm m}$ versus Ca$_{\rm m}$ (upper left panel), $S_{\rm H\alpha}$-$S_{\rm Ca}$ slope versus Ca$_{\rm m}$ (upper right panel), $S_{\rm H\alpha}$-$S_{\rm Ca}$ offset versus Ca$_{\rm m}$ (lower left panel), and average ff versus Ca$_{rm pl}$ (lower right panel). The retrieved values from the simulations are shown in black, and the observed values (stars in the 3600-3700 range) in red. 
}
\label{comp_simu_obs}
\end{figure*}

\begin{figure*}
\includegraphics{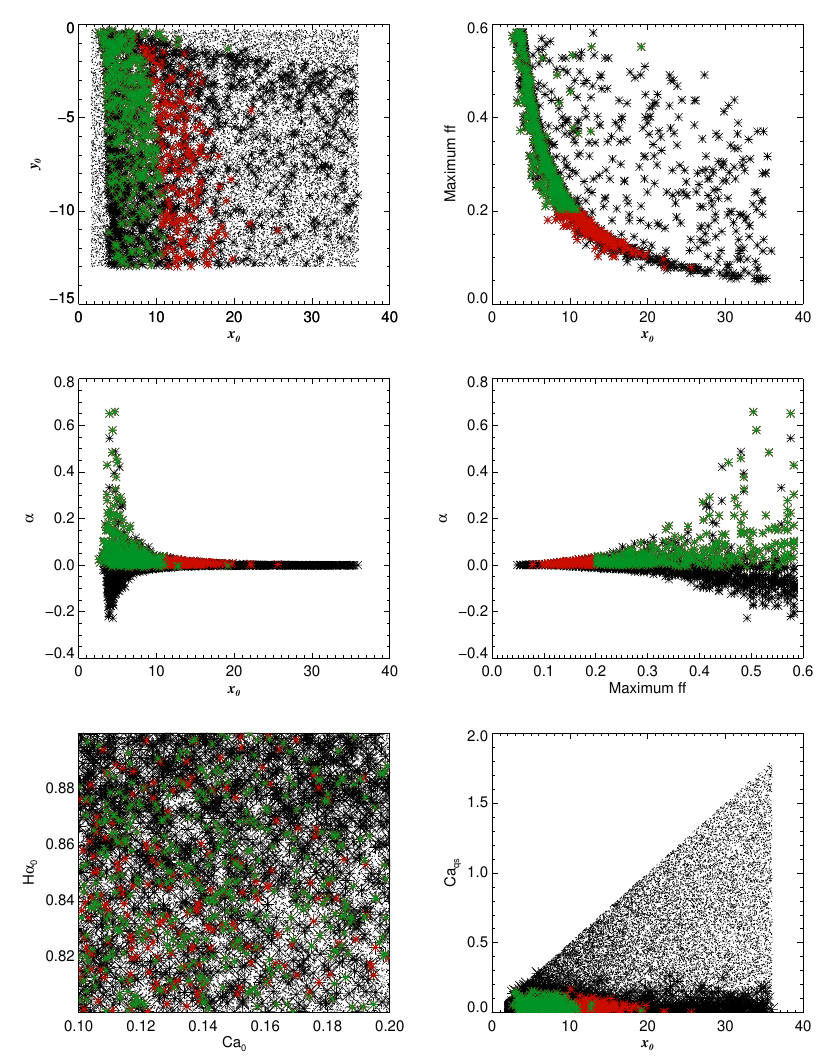}
\caption{Range of input and retrieved parameters from Test \#1: all tested parameters (dots), selection \#1 (stars), selection \#2 (coloured stars, red and green), and selection \#3 (red stars): $x_0$ versus $y_0$, maximum ff (over all stars) versus $x_0$, $\alpha$ versus $x_0$, $\alpha$ versus maximum ff, H$\alpha_0$ versus Ca$_0$, Ca$_{\rm qs}$ versus $x_0$. 
}
\label{simu_sel}
\end{figure*}

\subsection{Range of input parameters covered by the simulations}

We built a large amount of time series according to Eq.~\ref{eqtest1_ca} and Eq.~\ref{eqtest1_ha} to compare their properties with observations. We first consider 10$^4$ pairs of ($x_0$,$y_0$) values (related to $a$ and $b$ in Eq.~\ref{eqcram}), with $x_0$ chosen in the range [1.6;36] and $y_0$ in [-13;-0.56] respectively. We determined these limits based on conservative assumptions on the ff values of the most active stars in the sample as follows. 

\begin{itemize}
\item{We considered that the most active stars probably have a ff (in fraction of the disc) larger than 0.3 because \cite{houdebine10} found that active M dwarfs had ff of at least several times 0.1. Also, given the shape of our function, with H$\alpha_{\rm loc}$ equal to zero for 2$\times x_0$, we wish to be able to go up to at least 2.5$\times x_0$ to have at least a small positive slope regime between S$_{\rm H\alpha}(t)$ and S$_{\rm Ca}(t)$. Given the observed maximum Ca$_{\rm m}$ and H$\alpha_{\rm m}$ compared to the quiet star levels in the 3500-3900 K range (to be conservative), respectively around 9 and 1.6, this leads to $x_0$<36 and $y_0$>-13.}
\item{Furthermore, since we want to reach at least 2.5$\times x_0$, then a very large ff would be necessary to reach the observed Ca$_{\rm m}$ if $x_0$ is too small: since ff cannot exceed 1, this provides a lower limit for $x_0$. We consider here that for 2$x_0$, we have ff<0.6, leading to $x_0$>1.6. }
\item{Finally, given the lowest observed values of H$\alpha$ (0.17 compared to the quiet star level), if $y_0$ is too small, then the ff values required to reach this level would also be too high. We consider a ff threshold of 0.3 in this regime, leading to $y_0$<-0.56.}
\end{itemize}

Then, for each ($x_0$,$y_0$) simulation, a grid in ff (minimum and maximum) and Ca$_{\rm pl}$ is systematically explored. We chose ff$_{\rm min}$ and ff$_{\rm max}$ to vary between 0.1\% and 60\% and considered 50 regularly-sampled values of this parameter. 
We chose the 60\% threshold according to \cite{houdebine09}.
Ca$_{\rm pl}$ is chosen between a small value ($x_0$/20, chosen arbitrarily) and a maximum value which depends on ($x_0$,$y_0$): Each Ca$_{\rm pl}$ value corresponds to a straight line in the Ca-H$\alpha$ space (Fig.~\ref{ca_ha_schema}), and increasing Ca$_{\rm pl}$ corresponds to higher S$_{\rm H\alpha}$. We therefore chose the maximum Ca$_{\rm pl}$ so that all observed values (for stars in a given T$_{\rm eff}$ bin) are below the corresponding straight line.  
We considered 250 regularly-sampled values of Ca$_{\rm pl}$. 
Finally, Ca$_{\rm qs}$ is arbitrarily chosen to be small, randomly between 0 and $x_0$/20, and Ca$_0$ and H$\alpha_0$ are chosen randomly in the ranges [0.1-0.2 and [0.8-0.9] respectively, that is  close to the low activity regime of the observations. 

\subsection{Production of the time series}

For each of these time series, we add some noise (which is chosen equal to the typical uncertainty in our observations). This leads to a large set of Ca and H$\alpha$ time series, corresponding to a given pair of ($x_0$,$y_0$).

For each pair of time series, S$_{\rm Ca}(t)$ and S$_{\rm H\alpha}(t)$, which corresponds to a given Ca$_{\rm pl}$, ff$_{\rm min}$ and ff$_{\rm max}$ (and therefore an average ff) in the grid, we computed several quantities: Ca$_{\rm m}$ and H$\alpha_{\rm m}$, the correlation between the two time series, the slope and the offset from the linear fit S$_{\rm H\alpha}(t)$ versus S$_{\rm Ca}(t)$. Those quantities can then be compared with observations.

\subsection{Analysis of the time series}

For a given ($x_0$,$y_0$), we compared the simulations and the observations for the $N$ stars in a T$_{\rm eff}$ bin as follows. 
We first check if there is a complete overlap between observed properties and those in the simulation for each ($x_0$,$y_0$) pair when representing H$\alpha_{\rm m}$ versus Ca$_{\rm m}$, or if many stars are completely outside the range covered by the simulations for a given ($x_0$,$y_0$). If we are in the second case, this means that the considered ($x_0$,$y_0$) is not compatible with the observations in this simple model, otherwise they may be compatible. We illustrate this in Fig.~\ref{exsimu_ca_ha} for a given ($x_0$,$y_0$) pair verifying the condition, with a general shape corresponding to that shown in the lower panel in Fig.~\ref{ca_ha_schema}. This figure provides interesting insights on the impact of the local law in Eq.~\ref{eqcram} on the average values. The dispersion derived from the range covered by the simulations for the whole grid of parameters is indeed much larger than the observation. Since each region in this diagram corresponds to a different Ca$_{\rm pl}$ and ff regime, this suggests that there is an underlying relationship between  Ca$_{\rm pl}$ and ff. The colour code in the figure is related to different ff regimes. Note that in the very low activity regime, different ff values can be superposed on the figure (and therefore correspond to different Ca$_{\rm pl}$). In the high activity regime, there is a clear variation of the ff regime across the diagram. Superposed on this colour code, several lines corresponding to a fixed Ca$_{\rm pl}$ are also represented. Different ($x_0$,$y_0$) values will show some departure from this example but with similar trends. Depending on ($x_0$,$y_0$), the observed points will correspond to different regimes (in ff or Ca$_{\rm pl}$) and therefore to a different relationship between ff and Ca$_{\rm loc}$.

For each realisation of ($x_0$,$y_0$) verifying this condition, we identified the closest time series in this 2D space ($S_{\rm H\alpha}$ versus  $S_{\rm Ca}$) for each of the $N$ stars and retrieved the quantities computed for that time series as well as the corresponding Ca$_{\rm pl}$ and  ff. These retrieved quantities are then affected to the star. 
We computed a $\chi^2$ to evaluate the difference between the closest point to each observed star for the quantities corresponding to observations, summed over all stars. The $\chi^2$ are normalised by the observed uncertainties. 

In addition, we computed a few additional criteria. We performed a linear fit over the $N$ values of Ca$_{\rm pl}$ vs ff, whose slope is denoted $\alpha$, to evaluate if the two are correlated. 
We also performed a linear fit of the retrieved offset values versus the Ca$_{\rm m}$ (for Ca$_{\rm m}$<2) for the $N$ stars, whose slope can be compared with the observed one. We also considered the average and rms (over the $N$ retrieved values) in slope for Ca$_{\rm m}$ in the low activity regime. 

\subsection{Selection of realisations and results}

We discuss here the results for T$_{\rm eff}$ bin 3600-3700~K (27 stars). Results are similar for the 3500-3600~K bin.  
Out of 2 10$^4$ realisations of ($x_0$,$y_0$) pairs, 1912 are compatible according to the above criterion, so that all stars are included in the range covered by the simulations in Ca$_{\rm m}$ and H$\alpha_{\rm m}$, as illustrated in Fig.~\ref{exsimu_ca_ha}. 

We now analyse in more details this selection (hereafter selection \#1) of realisations. The $\chi^2$ corresponding to the Ca$_{\rm m}$ and H$\alpha_{\rm m}$ are usually low, with many values below 1. However, the $\chi^2$ corresponding to the slope is never excellent (minimum value of $\sim$4.5) and it is far worse for the offset. In addition, we never reproduce proper correlations between time series, as they are always close to 1 or -1 in this simple model, as in Sect.~\ref{sec51}. This means that the model lacks the complexity necessary to produce a diversity of complex time series that would allow to produce the proper correlations and slopes for each star. 

Two representative examples of the comparison are illustrated in Fig.~\ref{comp_simu_obs} for a given ($x_0$,$y_0$) pair corresponding to a reasonably good agreement. For each example, the upper left panel shows that according to this criteria, there is indeed a time series corresponding to each star in the H$\alpha_{\rm m}$-Ca$_{\rm m}$ space for that particular ($x_0$,$y_0$) pair. The upper right panel for each example shows the slope derived from those time series in comparison with the observed slope. For each star, there is a poor agreement, especially in the first example, leading to the fact that the offset of the linear fits (on the time series) does no exhibit the proper dependence on Ca$_{\rm m}$ (lower left panel). However, the dispersion in slope, large compared to the observations, is well reproduced in the second example. So even if the detailed behaviour of the observed stars is not retrieved because some ingredients are missing in the model, some properties can already be  reproduced. A  similar simulation but with Ca$_{\rm qs}$=0 leads to very different properties for the slope: at a given Ca level, they all have the same sign (as in Fig.~\ref{ca_ha_schema}). Therefore a small Ca$_{\rm qs}$ value different from 0, even if the same for all stars, is sufficient to introduce some degree of mixity in the slope signs in some of the realisations(but not all, as illustrated in the first example). This is true despite the fact that the best Ca$_{\rm qs}$ are small: the explored range is [0,$x_0$/20] but they are usually lower than $x_0$/100, as illustrated below. 

In a second step, we therefore selected all realisations of ($x_0$,$y_0$) for which the average and dispersion in slope in the low activity regime is close to the observed values:  The average of the $N$ retrieved slopes is selected between -0.1 and 0.1, and the average below 0.1. This leads to a selection of 591 realisations (hereafter selection \#2). We then refined this selection by adding the constraint that the largest ff (out of the $N$ stars) should be larger than 0.2 (hereafter selection  \#3, 401 stars). We study the properties of these selections in more details, as illustrated in Fig.~\ref{simu_sel}. The $x_0$ and $y_0$ values from selection \#1 covers a good fraction of the whole range (although we see that our minimum $x_0$ value is well chosen). Selection \#2 also covers a large range, however only small $x_0$ values are compatible with selection \#3. This is logical, because $x_0$ controls the necessary ff necessary to fit the observations (upper right panel). 

We now discuss an interesting property between the retrieved ff and Ca$_{\rm pl}$ values for each star which is illustrated in the lower right panel of each example in Fig.~\ref{comp_simu_obs}. This relationship is simply characterised here by the slope between Ca$_{\rm pl}$ and ff over the $N$ stars. When considering all realisations in selection \#1, both signs are present, with most realisations corresponding to negative $\alpha$ values for large maximum ff values (and small $x_0$). However, after selection of the stars with the best statistical properties regarding the slope (selection \#2) and ff range (selection \#3), the remaining realisations have mostly positive $\alpha$ values. 
Such a positive slope is something we might expect from the solar case for which such a  relation is observed, with a stronger Ca local chromospheric emission when the plages are larger, because they are associated to a larger local magnetic field \cite[][]{harvey99}. It is therefore possible that a similar relationship exists for M dwarfs. 
We note however that a linear fit on the retrieved Ca$_{\rm pl}$  versus ff does not go through zero, and corresponds to Ca$_{\rm pl}$ values for ff=0 that are typically between a few units and 20. It does not means that there is no small structures with a weak emission, but this simple model is not sensitive to those if they exist. 
The shape of the observed H$\alpha_{\rm m}$ versus Ca$_{\rm m}$  makes it very difficult to reach the very low Ca$_{\rm pl}$ regime in Fig.~\ref{exsimu_ca_ha}.

\end{appendix}

\end{document}